\documentclass[10pt,journal]{IEEEtran} 

\usepackage{multirow}
\usepackage{xcolor, colortbl}
\usepackage[numbers]{natbib}
\usepackage{color,soul}
\usepackage{multirow}
\usepackage{varwidth}
\usepackage{xcolor, colortbl}
\usepackage[pdftex]{graphicx}
\usepackage{graphicx}

\begin{document}

\title{A Survey on Threat Situation Awareness Systems: Framework, Techniques, and Insights}

\author{Hooman Alavizadeh, Julian Jang-Jaccard, Simon Yusuf Enoch, Harith Al-Sahaf, Ian Welch, Seyit A. Camtepe, Dong Seong Kim
\thanks{Hooman Alavizadeh is with The Massey University, Auckland, New Zealand, email: h.alavizadeh@massey.ac.nz}
\thanks{Julian Jang-Jaccard is with The Massey University, Auckland, New Zealand, email: J.Jang-jaccard@massey.ac.nz}
\thanks{Simon Yusuf Enoch is with The University of Queensland, Australia, email: sey19@uclive.ac.nz}
\thanks{Dong Seong Kim is with The University of Queensland, Australia, email: dan.kim@uq.edu.au}
\thanks{Harith Al-Sahaf and Ian Welch are with The Victoria University of Wellington (VUW), email: \{harith.al-sahaf, Ian.Welch\}@vuw.ac.nz}
\thanks{Seyit A. Camtepe is with CSIRO Data61, Australia, email: Seyit.Camtepe@data61.csiro.au}
\thanks{Dong Seong Kim is with The University of Queensland, Australia, email: dan.kim@uq.edu.au}
}

\maketitle  

\begin{abstract}
Cyberspace is full of uncertainty in terms of advanced and sophisticated cyber threats which are equipped with novel approaches to learn the system and propagate themselves, such as AI-powered threats. To debilitate these types of threats, a modern and intelligent Cyber Situation Awareness (SA) system need to be developed which has the ability of monitoring and capturing various types of threats, analyzing and devising a plan to avoid further attacks. This paper provides a comprehensive study on the current state-of-the-art in the cyber SA to discuss the following aspects of SA: key design principles, framework, classifications, data collection, and analysis of the techniques, and evaluation methods. Lastly, we highlight misconceptions, insights and limitations of this study and suggest some future work directions to address the limitations.
\end{abstract}





\maketitle

\section{Introduction} \label{sec:intro}
\subsection{Motivation} \label{subsec:motivation}
Situational Awareness (SA) was firstly introduced in a comprehensive work by Endsley in~\cite{endsley1988design}. Based on Endsley definition, an SA system consists of three main component which are \textit{Perception}, \textit{Comprehension}, and \textit{Projection}. This work is considered as the main reference model for SA research and has been widely expanded and applied in various range of research contexts. For instance, the application of SA on aircraft was presented in~\cite{endsley1995measurement} which studied the SA in aircraft pilots aiming to increase the likelihood of finding the optimal decisions in a complex real-time situations. When it comes to the Cyber security domain, Cyber SA may be defined as the preparation, incorporation, processing, and evaluation of data related to a given system to understand the system's environment to be able to predict and respond accurately to potential Cyber threats against the given system or network~\cite{barford2010cyber,albanese2017computer}. 
Situation Awareness in cyber space consists of three seminal aspects~\cite{barford2010cyber}: ($i$) Situation Recognition (also called as situation perception) deals with identifying the occurrence of an attack alongside with the type, source, and target of the attack. This aspect involves awareness of the collected data and information in terms of quality, truthfulness, completeness, and freshness. ($ii$) Situation Comprehension including attack impact assessment (damage assessment) for both current and future impacts. This aspect also involves awareness of attacker's behavior which considers attack's trend and intent. Situation Comprehension aspect needs to know about the cause of current situation. ($iii$) Situation Projection including awareness of how the situation evolve and may have further affects.
A comprehensive design of SA system would help the decision makers to be aware of the current situation and the security posture of the system and increase their understanding of the situation up to the decision point. The planning and execution (of the response actions) occur once the decision is made based on the current situation.

However, most of the current approaches in the literature to gain cyber SA focus on the lower and abstract levels of SA techniques such as vulnerability analysis which may use attack graphs (AGs), alert correlation and intrusion detection techniques~\cite{kim2016long}, analyzing attack trend~\cite{mansmann2012streamsqueeze}, information flow and taint analysis~\cite{lakkaraju2004nvisionip}, causality analysis and forensics, damage assessment~\cite{liu2010cross}. However, higher SA level ranging from SA perception to projection are still missing and performed manually by experts which is time consuming and error-prone. There still a lack on designing a SA systems which be able to react to a dynamic environment by the ability to adapt itself without the high and intensive interaction with human or agents.

\subsection{Major Commercial Perspective} \label{subsec:Commercial}
Gartner\footnote{https://www.gartner.com/en/documents/3945589} raised a concern over the sheer number of alerts generated by currently available market level threat monitoring systems (such as IDPS). To reduce the effort required by security analysts having to deal with isolated alerts, the use of artificial intelligence techniques to group various alerts together to create a single incident or to describe a chain of related activities has been emerged. Gartner also suggest a requirement for deploying additional threat monitoring sensors inside the network to detect threats that have bypassed traditional controls (e.g., firewalls). Further Gartner reports the move of many IDPS vendors deploying their services into public cloud (i.e., IaaS) environment than organization network firewall solutions making the cloud-based monitoring capability even more critical. The shift is caused by two reasons; (1) more organizations tend to move their high-value data and services to the cloud, and (2) take advantage of extra layer of protection provided by cloud vendors.

The report by McAfee\footnote{https://www.mcafee.com/blogs/other-blogs/mcafee-labs/mcafee-labs-2020-threats-predictions-report/} stats the potential raise of less-skilled adversaries to have more access and broader capabilities to create and weaponize deepfake (e.g., use of deep learning with adversarial effect) content. Similarly, it predicts that adversaries will use artificial intelligence to produce extremely realistic text, images, and videos capable of bypassing many biometric-based user authentication mechanisms. The report also illustrates the concern over the vulnerabilities in Application Programming Interfaces (API) which expose to public to allow the access into organization software platforms and app ecosystems. The report shows that attackers tend to shift their attack path from web app to the API as a new attack entry point.


Symantec\footnote{https://docs.broadcom.com/doc/istr-24-2019-en} share the latest insight into global threat activity, cyber attacker trends, and attack motivations. The annual report warns the increasing attack through formjacking where attackers load malicious code onto retailer's websites to steal shoppers' credit card details, reporting close to 5,000 unique websites were compromised on average every month including Ticketmaster and British Airways. Multi-faceted attacks that combines multiple attack techniques, such as combining link as a smoke screen, cryptojacking, phishing, into a single attack is on the rise to avoid detection thus a detection technique designed to stop a single type of attack is increasingly insufficient. More targeted attack (e.g., spear phishing) has also increased to infiltrate organizations while using intelligence to gather as much information about the target organization.

\subsection{Key Design Principles} \label{subsec:SA_decisions}
The fundamental design principle for developing SA systems lies in the understanding of multiple facets of the cyber landscape through the following key concerns: 

\begin{itemize}
\item[] {\bf 1-- What is happening}: `What is happening' in SA refers to detecting whether is any ongoing attack in the system's environment, or what resources have been compromised? This also includes the impact of an attack. However, this is a part of perception in SA, and mostly involves automated data gathering tools and pre-process of huge amount of gathered data. The quantity and quality of gathered data such as Intrusion Detection System (IDS) and Firewall logs, vulnerability scanning tools, anti-malware log files, etc. determine how effective the SA can answer this question.

\item[] {\bf 2-- Why is it happening}: `Why is it happening' in SA refers to monitoring the system's environment using the vulnerabilities, security holes, security alerts, to be aware of the potential threats and attacks. Moreover, this item refers to the way in which the situation is evolving, including attack tracking, attack behaviors and strategies analysis. In this stage, more reasoning and analysis techniques need to be heavily used.


\item[] {\bf 3-- What may happen in future}: `What may happen in future' indicates the ability of forecasting possible futures, along with the probabilities and anticipate damage potential. It includes current situation knowledge and the possibility of its evolution alongside with knowledge about the behaviors of the adversaries. This question is a part of SA projection which answers the questions such as, what situations may be possible based on current system components, security posture and threats. Moreover, what possible ways are for further evolving and exploiting current situations be exploited?

\end{itemize}

\vspace{1mm}

However, the main focus of this paper is to answer these seminal questions based on the current state-of-the-art which are related to SA monitoring. Answering these questions can further be used by reaction and response phases which mainly are planning, response, and prediction. The question of whether an appropriate reaction or response can be satisfactorily done is greatly dependent upon the SA capability to deal with concerns 1--3.

\subsection{{Review of Existing SA Survey Papers }}
Some efforts have been made to understand the state-of-the-art SA systems in cyber security realm. In here, we compare our survey paper with the existing SA survey papers focusing on comparing the key contributions, design, and application.

We compared our survey paper and the existing survey papers in terms of their contributions, key design, and classification together with the following principal questions:
\begin{itemize}
\item[] {\bf Q1}: What are the main threats and attack behaviors in a cyber space?
\item[] {\bf Q2}: What AI techniques are more commonly explored by AI-powered attacks that need to be captured in Cyber situation awareness design.
\item[] {\bf Q3}: How a comprehensive data collection need to be performed to be able to capture most threats?
\item[] {\bf Q4}: What are the data types and how they can be classified to be used for situation monitoring context?
\item[] {\bf Q5}: What theoretical and empirical techniques have been used in literature to design and develop situation awareness systems?
\item[] {\bf Q6}: How to map the situation awareness principles including situation perception, comprehension, and projection to a related situation from low level of understanding to high perception level in a framework such as data gathering, analysis, and gaining high awareness?
\item[] {\bf Q7}: What are commonly used tools and prototypes that can be used in analysis of situation awareness?
\item[] {\bf Q8}: What are the specific limitations in situation awareness system in each level?
\end{itemize}


\citet{franke2014cyber} studied the systematic review of the scientific literature on cyber situational awareness. They reviewed and clustered 102 articles in SA context. Although their categorization and mapping of the reviewed studies based on their area of focus are well-studied, they did not extensively discussed the main techniques and methods on each article in terms of design, analyze, and development of situation awareness system.

\citet{leau2015network} surveyed the network security situation forecasting techniques. They categorized the techniques into three categories: machine learning, Markov models, and Grey theory. The authors explained each technique in details under these categories, which is useful as a reader can understand the fundamentals of each technique. However, they did not provide the enough discussion on more possible techniques for different attack behaviours. Moreover, their research can not enable the readers to find out the answers to the principal questions Q4--Q8.

\citet{husak2018survey} published a survey paper on situation awareness focusing on prediction, and forecasting methods used in cyber security. They expensively reviewed and categorized the methods and techniques based on (i) discrete models (such as attack graphs, and Markov
models, and Bayesian networks), and (ii) continuous models (such as time series and grey models), and (iii) machine learning and data mining approaches. However, they mainly discussed situation projection level and the basic and critical concepts of situation awareness such data collection and pre-processing were missing in their paper. Although this classification covers a large portion of literature, it misses other perspectives of SA such as implementations and tools. However, from the reader's perspective, it is challenging to understand the principal questions Q1--Q3 and Q7. 

\subsection{Key Contributions \& Scope} \label{subsec:scope_contribution}
The main contributions of this survey paper have been highlighted as follows: 

\begin{itemize}
\item[--]  We extensively surveyed the Situational Awareness (SA) frameworks and classified them based on three main parts: data gathering, techniques and analysis, situation awareness and visualization.
\item[--] We surveyed the most commonly used approaches, techniques, and methodologies used in the existing literature to develop SA systems, which embrace the theoretical backgrounds  of anomaly analysis, Artificial Intelligence (AI), Game Theory, Machine Learning (ML), and so forth, and highlight the limitations.
\item[--] We discussed the various application prototypes and tools which have been used or applied to SA techniques and systems.
\item[--] {We discussed misconceptions, insights, and limitations obtained from this extensive survey.}
\end{itemize}

\subsection{Paper Structure}
This paper is organized as follows:
\begin{itemize}
\item[--] Section~\ref{sec:threats} explains the attack and threat behaviors including advanced and AI-based threats discussed by existing SA studies. 
\item[--] Section~\ref{Data-Gathering} surveys the main data gathering approaches for SA systems. This section also includes explanation of data types and data sources used for SA monitoring. Further, we classify those types of data based on the different criteria such as availability, accessibility, complexity of use, and usability for SA system.

\item[--] Section~\ref{Analysis} provides a comprehensive review and classification of the existing techniques and approaches used to analyze cyber SA in various systems and contexts, along with the discussions of main limitations of each technique.

\item[--] Section~\ref{sec:SA} discusses the main situation awareness phases including threat evaluation, decision making, and planning alongside with visualization leading to high-level understanding of the system situation in the projection level.

\item[--] Section~\ref{sec:discussion} discusses the insights and lessons learned from our study and suggests future research directions.
\end{itemize}



\section{Main Threats and Attack Behaviours} \label{sec:threats}
As ICT continues to evolve, so does threats and attacks incidences.
 In this section, we survey the existing literature on threats and attacks considered in SA systems. In addition, we survey the state-of-the-art techniques on AI-based cyber-attacks, and the techniques of monitoring such attacks. 
Specifically, we use the MITRE ATT\&CK model that groups attacks based on adversary tactics and techniques to describe the characteristics of the attacks. In addition, we use Microsoft's STRIDE threat model \cite{LeBlanc:STRIDE} to map the attacks with their corresponding threats. The STRIDE threat model captures the unique characteristics of attacks that pose a particular type of threat.

\subsection{Threat and Attack Landscape}
\label{sec:attack_landscape}
The STRIDE threats model categorized threats into six categories ``Spoofing, Tampering, Repudiation, Information Disclosure, Denial of Service, and Elevation of privilege''.
{\em Spoofing} specifies when an adversary disguises by falsifying data or information to gain an illegitimate advantage. {\em Tampering} describes when an adversary modifies components to caused operations disruption. {\em Repudiation} is a threat that identifies when an adversary rejects actions because the actions cannot be properly tracked. {\em Information disclosure} is a threat that specifies a leak confidential information to the people who are not supposed to see it. The next STRIDE threat is {\em Denial of Service}, and it specifies when valid users are denied resources as a result of the adversary by means of exploiting the system's vulnerabilities (e.g., memory, bandwidth, etc). The last STRIDE threat is the Elevation of Privilege and it specifies when an adversary gains unauthorized privileges by exploiting the system's weaknesses.


The ATT\&CK group attacks into the following tactics: Initial Access,	Execution,	Persistence,	Privilege Escalation,	Defense Evasion, Credential Access,	Discovery,	Lateral Movement, Collection, Command and Control, Exfiltration, and Impact. We describe the attack categories as follows, then we provide a summary of the attack techniques used in SA systems in Table \ref{tbl:attacks} (grouped according to attack tactics).

\begin{itemize}
\item[--] {\em Initial Access}: An adversary can use different vectors to gain entry and foothold into a system. The initial access is the attack technique that allows the adversary to have initial entry to a system. This type of early-stage attack has been considered in many SA research. Brancik and Ghinita~\cite{brancik2011optimization} considered an insider attack where malicious software is planted on the organization's computer to allow access to other machines remotely. The malware is expected to affect certain files before the adversary stage an attack. As a result of the initial access, other forms of attack tactics such as Execution, Exfiltration, \textit{etc} may happen.

\item[--] {\em Execution}: Execution consists of injecting adversary-controlled code into a program, remotely or locally. Execution is often linked to other attack tactics. For example, the work in \cite{dutt2013cyber} presented modeling and detection of attacks for the SA system where an adversary is able to execute a script or command with a service control manager as part of the {\em lateral movement}. 

\item[--] {\em Persistence}: An adversary access can be lost as a result of changes to the system (e.g., password changed, system restart, \textit{etc}). The persistence attack consists of the techniques used by the adversary to continually maintain a foothold on the system. One example of the persistence techniques is presented by He \textit{et al.} \cite{he2018scpn} for a cyber SA system in the IoT network. In the work, He \textit{et al.} considered an adversary who is able to expand persistence by using malware. Specifically, the attacker is able to turn on persistence root privilege of a TV in a smart home via an Android tablet portal.

\item[--] {\em Privilege Escalation}: An adversary can exploit system vulnerabilities to gain permissions to a user account, or even gain higher privileges than the normal user account. The adversary can then utilize the newly gain account privileges to potentially perform damages on the system. In privilege escalation, an adversary may steal normal user account credentials to have initial access before escalating the user privilege to root privilege to gain full control of the system or perform a lateral movement in network \cite{zhang2018network,yang2014attack,kou2019research,he2018scpn}.

\begin{table*}
\centering
\scriptsize
\caption{A review of different attack techniques and threats used in existing SA systems}
\vspace{-3mm}
\label{tbl:attacks}
\begin{tabular}{|l|l|p{1.13cm}|l|l|l|}
\hline
\multirow{2}{*}{\begin{tabular}[c]{@{}l@{}}\textbf{Attack Tactics}\\ (ATT\&CK)\end{tabular}} & \multicolumn{5}{c}{\cellcolor{gray!35!white} \textbf{THREATS}} \\  \cline{2-6} 
 & \cellcolor{gray!30!white}\textbf{Spoofing} & \cellcolor{gray!30!white}\textbf{Tampering} &  \cellcolor{gray!30!white}\textbf{Information Disclosure} & \cellcolor{gray!30!white}\textbf{Denial of Service} & \cellcolor{gray!30!white}\textbf{Elevation of Privileges} \\ \hline
\cellcolor{gray!30!white}Initial Access & \begin{tabular}[c]{@{}l@{}}trusted relationship \cite{brancik2011optimization}\\ valid account\cite{trieu2018artificial}\\ Fake event  \cite{lu2018network}\end{tabular} & \cellcolor{gray!10!white} & spear-phishing \cite{gouglidis2016threat} & \cellcolor{gray!10!white} & \cellcolor{gray!10!white} \\ \hline
\cellcolor{gray!30!white}Execution &  \cellcolor{gray!10!white} & \cellcolor{gray!10!white} & \cellcolor{gray!10!white} & exploit execution \cite{gouglidis2016threat} & code execution \cite{rajivan2017impact}  \\ \hline
\cellcolor{gray!30!white}Privilege Escalation & \cellcolor{gray!10!white} & unauthorized write \cite{zhang2018network} & \cellcolor{gray!10!white}  & Hijack/DoS \cite{zhang2018network} & \begin{tabular}[c]{@{}l@{}}Elevation of privileges \cite{yang2014attack}\\ Root access\cite{kou2019research}\\ User privileges\cite{zhang2018network}\\ Root privileges \cite{zhang2018network}\end{tabular} \\ \hline
\cellcolor{gray!30!white}Defense Evasion & Backdoor \cite{trieu2018artificial} & \begin{tabular}[c]{@{}l@{}}Authentication\\ bypass \cite{kou2019research}\end{tabular}  & unauthorized access \cite{zhang2018network} & \cellcolor{gray!10!white} & \cellcolor{gray!10!white} \\ \hline
\cellcolor{gray!30!white}Persistence & \cellcolor{gray!10!white}  & \cellcolor{gray!10!white} &\cellcolor{gray!10!white}  &\cellcolor{gray!10!white}  & Extended privileges\cite{he2018scpn} \\ \hline
\cellcolor{gray!30!white}Credential Access & Phishing \cite{vinayakumar2019deep} & Defacement \cite{trieu2018artificial}  & \begin{tabular}[c]{@{}l@{}}credential leak\cite{Vinayakumar2018}\\ Password brute force \cite{trieu2018artificial}\end{tabular} & \cellcolor{gray!10!white} & \cellcolor{gray!10!white} \\ \hline
\cellcolor{gray!30!white} Discovery & \begin{tabular}[c]{@{}l@{}}Probing \& Mscan \cite{wu2016big}\\ Abnormal scan \cite{goodall2018situ}\\ IP sweep \cite{kou2019research}\\ Connections Discovery  \cite{zhang2018network}\end{tabular} & \cellcolor{gray!10!white}  & Reconnaissance \cite{ioannou2019markov} & System Service\cite{wu2016big} &  \cellcolor{gray!10!white}\\ \hline
\cellcolor{gray!30!white}Lateral Movement & \begin{tabular}[c]{@{}l@{}}Internal  spearphishing \cite{yang2014attack}\end{tabular} & \cellcolor{gray!10!white} & Internal  spearphishing \cite{yang2014attack} & \cellcolor{gray!10!white} & \begin{tabular}[c]{@{}l@{}}sequential attack \cite{dutt2013cyber}\\ Multi-step attacks \cite{he2018scpn}\end{tabular} \\ \hline
\cellcolor{gray!30!white}Collection & \cellcolor{gray!10!white}  & \cellcolor{gray!10!white} & \begin{tabular}[c]{@{}l@{}}Information collection \cite{ioannou2019markov}\\ information leakage \cite{kou2019research}\end{tabular} & \cellcolor{gray!10!white} & \cellcolor{gray!10!white} \\ \hline
\cellcolor{gray!30!white}Command and Control & \cellcolor{gray!10!white} & \cellcolor{gray!10!white} & \cellcolor{gray!10!white} & \begin{tabular}[c]{@{}l@{}}run command \\ and control for DoS \cite{guo2019ddos}\end{tabular} & have full control \cite{rajivan2017impact} \\ \hline
\cellcolor{gray!30!white}Exfiltration &  & move and install false data\cite{lu2018network}  & data exfiltration \cite{goodall2018situ} & \cellcolor{gray!10!white} & \cellcolor{gray!10!white} \\ \hline
\cellcolor{gray!30!white}Impact & \cellcolor{gray!10!white} & Modify files\cite{he2018scpn} & \cellcolor{gray!10!white} & \begin{tabular}[c]{@{}l@{}}DoS \cite{Vinayakumar2018,yang2014attack}, \\ Smurf \& Mailbomb \cite{wu2016big}\\ DDoS \cite{guo2019ddos}\end{tabular} & \cellcolor{gray!10!white} \\ \hline
\end{tabular}
\end{table*}

\item[--] {\em Defense Evasion}: are the techniques used by an adversary to avoid detection during attacks. As a result of this technique, an adversary can get full trust in the targeted system. For instance, an adversary may take advantage of vulnerable components in a web application to bypass security rules to access a database. In addition, it may also give the adversary the ability to remotely run system commands and install malware. Once installed, detection is difficult as data will be highly obfuscated. Other example includes disabling security tool, deleting registry, \textit{etc} \cite{trieu2018artificial,kou2019research,zhang2018network}.

\item[--] {\em Credential Access}: Credential access are the techniques used by an adversary to steal account and passwords credentials to achieve their objectives. Various attack techniques such as phishing, man-in-the-middle, brute force, network sniffing, \textit{etc} may be employed to leak or steal credentials from a company. This type of attack technique used in SA systems has been studies in \cite{Vinayakumar2018,trieu2018artificial}.

\item[--] {\em Discovery}: An adversary may perform a pre-attack passive information gathering before getting into a system. By doing so, the adversary can gain knowledge of the target system, including entry points and how to achieve his objectives. In Table \ref{tbl:attacks}, we show examples of discovery techniques used in the existing SA systems. For example, Wu \textit{et al.} \cite{wu2016big} proposed a SA mechanism based on the analysis of big data in the smart grid. In the system, they considered IP sweep and Mscan attack techniques to gain knowledge of the active devices.

\item[--] {\em Lateral Movement}: An adversary can exploit a sequence of systems and accounts to reach their objective. For example, in a three-tier system, an adversary will have to perform a multi-stage attack across multiple systems and account to compromise a target in the last hierarchy. For instance, the work in \cite{dutt2013cyber,he2018scpn} has considered attack scenarios for SA systems where the adversary combines different steps sequentially to launch an attack on a specific target, where the outcome of one step serves as an input to its subsequent steps.

\item[--] {\em Collection}:  An adversary may use various-compromise information gathering techniques before stealing data on the target system, such techniques include automated internal data collection, user email data collection, \textit{etc}. Ioannou \textit{et al.} \cite{ioannou2019markov} showed an example adversary collection technique based on sensitive data from a database residing on a system prior to exfiltration. Specifically, the adversary collects and uploads data to an adversary's server during exfiltration via specially crafted malware.

\item[--] {\em Command and Control}: Adversary may establish command and control to plan, direct normal traffic to control their target. for instance, an adversary may communicate with a commonly used open port as a means of relaying commands and controlling compromised systems.

\item[--] {\em Exfiltration}: involves the techniques that an adversary employs to move or copy authorized data from a system. Lu \& Feng \cite{lu2018network} developed a cyber SA framework for an industrial control system for an attack, where the entire system may have its integrity compromised by having unauthorized commands injected into systems to create fake events then move data to the adversary. 

\item[--] {\em Impact}: An attacker can impact system availability and integrity by manipulating or destroying its operations using various techniques. Here, the adversary can change the normal data route to a database in order to provide cover for a confidentiality breach. The techniques used can also include the tampering of data \cite{he2018scpn}.
\end{itemize}

\subsection{Advanced Attacks} \label{subsec:charac-advanced-attacks}


Adaptive (advanced) attacks are known as intelligent attacks. In these types of attacks, the attackers are adaptive to dynamically changing system conditions and external environmental conditions, as they take into consideration both physical and cyber accessibility. These attackers also have intelligence with regard to their resources, executing adaptive attacks~\cite{tramer2020adaptive} that wisely manage their resource limits and at the same time opportunistically seek to compromise an entire system.
\citet{kaloudi2020ai} investigated the AI-powered cyber attacks and mapped them onto a proposed framework with new threats including the classification of several aspects of malicious using AI during the cyber-attack life cycle. In the following, we discuss the advanced cyber-attacks from existing literature. AI offers significant benefits in terms of innovation and automation in different domains. However, cyber-criminals utilizes the AI technologies to improve their attack strategies in conjunction with other conventional attack techniques discussed in Section \ref{sec:attack_landscape}. 
It is important for SA systems to take into account AI-powered attacks as well as the methods to mitigate them. However, there are only a few works that attempt to develop SA systems taking into account AI attacks. Jiang \cite{jiang2020improving} developed an approach to improve SA representation and learning using collective AI (path-based embedding and graph neural network) over knowledge graphs. Specifically, the author introduced four ideas for prediction with collective AI for SA; prediction ensemble, data aggregation, representation aggregation, and joint representation learning. 
Here, we survey the state-of-the-art techniques on AI-based cyber-attacks and then go further to map them to the techniques used in monitoring and defending them. As a result, this will provide decision-makers with insight into AI-based attacks, their detection, and mitigation approach. \\
Attackers can utilize AI techniques to achieve attacks and in other cases, the attackers can exploit weaknesses of the AI-based techniques to successfully compromise a system. We categorized the attacks based on {\em AI-supported} and {\em adversarial} attacks. We summarize them in Table \ref{tbl:AI_attacks} and discuss them as follows.

\begin{table*}
\scriptsize
\caption{AI-based cyber attacks, their techniques and mitigation}
\vspace{-3mm}
\label{tbl:AI_attacks}
\begin{tabular}{|l|l|l|l|l|}
\hline
Category & Paper & Attack type & Technique & Detection/Mitigation \\ \hline
\multirow{9}{*}{\begin{tabular}[c]{@{}l@{}}AI-supported\\ attacks\end{tabular}} & \cite{seymour2016weaponizing} & \begin{tabular}[c]{@{}l@{}}Target discovery, Automated \\spear phishing\end{tabular} & \begin{tabular}[c]{@{}l@{}}Long short-term memor, and Markov \\chains (deep learning)\end{tabular} & \begin{tabular}[c]{@{}l@{}}Using a detection system by incorporating the new systetic\\ URLs\end{tabular} \\ \cline{2-5} 
 & \cite{hitaj2019passgan} & Credential access & Deep learning and GANs & Defenses have not implemented yet \\ \cline{2-5} 
 & \cite{trieu2018artificial} & Credential access & ML algorithm (Torch-rnn) & \begin{tabular}[c]{@{}l@{}}Use AI-based password brute force algorithms  to prevent \\ users from choosing poor new passwords\end{tabular} \\ \cline{2-5} 
 & \cite{stoecklin2018deeplocker} & Deep Locker & Deep neural networks & No defense is described \\ \cline{2-5} 
 & \cite{xia2019genpass} & Credential access & \begin{tabular}[c]{@{}l@{}}Probabilistic context-free grammars \cite{weir2009password} \\ and recurrent neural network  model\end{tabular} & \begin{tabular}[c]{@{}l@{}} Dynamic personalized password policy based on user's\\ personality traits \cite{guo2020nudging},  interpretable probabilistic password\\ strength meters via deep learning \cite{pasquini2020interpretable}\end{tabular} \\ \cline{2-5} 
 & \cite{pal2019beyond} & \begin{tabular}[c]{@{}l@{}}Credential access\\ (credential tweaking)\end{tabular} & \begin{tabular}[c]{@{}l@{}}A generative model-based  on sequence-to-\\ sequence learning, A discriminative model \\ based on word embedding techniques\end{tabular} & \begin{tabular}[c]{@{}l@{}}personalized password strength meters using NN based \\ word embedding techniques, password strength meters\end{tabular} \\ \cline{2-5} 
 & \begin{tabular}[c]{@{}l@{}}\cite{bahnsen2018deepphish}\\ \cite{seymour2016weaponizing}\end{tabular} & \begin{tabular}[c]{@{}l@{}}bypass AI phishing \\ detection systems\end{tabular} & Deep Neural Networks & \begin{tabular}[c]{@{}l@{}}Detection system by incorporating the new systemic URLs\end{tabular} \\ \cline{2-5} 
 & \cite{kirat2018deeplocker} & \begin{tabular}[c]{@{}l@{}}concealment against antivirus, \\ disable countermeasures,\\ Unlocks malicious payload\end{tabular} & Deep Neural Networks & \begin{tabular}[c]{@{}l@{}}No defense is implemented, however, the authors in  \cite{kirat2018deeplocker} \\ recommended host-based monitoring, AI usage monitoring, \\AI lock picking, etc\end{tabular} \\ \cline{2-5} 
 & \cite{yao2017automated} & Crowdturfing Attacks & Recurrent Neural Networks & \begin{tabular}[c]{@{}l@{}}Using lossy transformation  introduced by the RNN training \\and generation cycle.\end{tabular} \\ \hline
\multirow{5}{*}{\begin{tabular}[c]{@{}l@{}}Adversarial\\ attacks\end{tabular}} & \cite{hitaj2017deep} & Evasion/deception & GANs & \begin{tabular}[c]{@{}l@{}}No defense is implemented  but \cite{hitaj2017deep} have recommended: \\fully homomorphic encryption \cite{sun2018private},  privacy-preserving \\collaborative learning \cite{shokri2015privacy}, differential privacy  \cite{abadi2016deep}  \\at different granularities.\end{tabular} \\ \cline{2-5} 
 & \cite{ateniese2015hacking} & information leak & Meta-classifier &  \\ \cline{2-5} 
 & \cite{papernot2016limitations} & Deception against  DNNs & Crafting algorithm & improve the training phase  (e.g., multilayer feedback \cite{hornik1989multilayer} ) \\ \cline{2-5} 
 & \cite{tramer2016stealing} & \begin{tabular}[c]{@{}l@{}}model extraction attack\end{tabular} & \begin{tabular}[c]{@{}l@{}}Generic equation solving  attack for models \\ with a logistic output layer.Path-finding \\ algorithm  with decision trees\end{tabular} & \begin{tabular}[c]{@{}l@{}}Rounding confidences, differential privacy, ensemble \\methods \cite{tramer2016stealing}\end{tabular} \\ \cline{2-5} 
 & \cite{hu2017generating} & Defense evasion & GAN based algorithm & \begin{tabular}[c]{@{}l@{}}autoencoders to map adversarial  samples to clean  input data\\ defensive distillation \cite{papernot2016distillation}\end{tabular} \\ \hline
\end{tabular}
\end{table*}

\subsubsection{ AI-based attacks}
In this section, we discuss attacks supported by AI, where advanced technologies are leveraged to power cyberattacks.
Hitaj \textit{et al} \cite{hitaj2019passgan} presented an approach to generate high-quality password guesses by automatically learning the distribution of real passwords from actual password leaks using deep learning and Generative Adversarial Networks (GANs). Their results showed that their approach surpasses rule-based and other ML password guessing tools, even without any a-prior knowledge on the passwords. Seymour \& Tully \cite{seymour2016weaponizing} presented a program named Network Automated Phishing and Reconnaissance which is designed to exploit social media users based on two forms of deep learning models; long short-term memory (LTSM) and Markov chains. Specifically, the program targets vulnerable twitter users that are more vulnerable to social engineering attacks by analyzing their tweets and then based on their tweets it creates relevant replies with a shortened obfuscated link to achieve a phishing attack on the target.
Stoecklin \cite{stoecklin2018deeplocker} presented an AI-powered evasive malware tool named Deep Locker that is able to conceal itself in other applications until it is unlocked based on certain trigger conditions such as geo-location, facial recognition, software, and user activity. It is difficult for defenders to determine the pattern of the deep locker and hence, it is complex to develop defense countermeasures for them \cite{kirat2018deeplocker}.
Rhodes \cite{Rhodes:Strategic_comments} provided an approach to infect multiple systems using automated self malware propagation. The author also described that this propagation can be achieved a large number of
systems while avoiding possible detection. Besides, other research work showed that AI-supported attacks can achieve faster and efficient attacks \cite{Batt2019}.

\subsubsection {Adversarial attacks}
In this section, we discuss attacks on ML models.
Hitaj \textit{et al.} \cite{hitaj2017deep} implemented an inference attacks on deep neural networks in a collaborative setting. In particular, they showed how an attacker can exploit the real-time nature of the learning process, where an attacker deceives a victim into releasing more accurate information on sensitive data. In their work, they showed that a distributed, federated, or decentralized deep learning approach can be attacked and thus, cannot protect the participant information (from the training sets).
Ateniese \textit{et al.} \cite{ateniese2015hacking} presented an approach to attack ML classifiers and other statistical information that can be revealed from them. The authors developed a meta-classifier that is trained to attack other ML learning classifiers to retrieve sensitive information
or patterns from the training set. The authors demonstrated how their approach could obtain unauthorized participants' information from trained voice recognition systems, that are not captured by privacy-preserving models or differential privacy.
Papernot \textit{et al.} \cite{papernot2016limitations} showed how an adversary can manipulate input fed to deep learning models that are used to infer and reveal the identity of object/persons in a blurred image. The authors showed that an adversary can make the model misclassify the input, thus producing incorrect outputs. 
Tram{\`e}r \textit{et al.} \cite{tramer2016stealing} presented attacks against online services of Amazon and BigML, where the work showed that it is possible for an adversary with black-box access, but no prior knowledge of an ML model's parameters or training data to steal ML models that are based on only predictions on input feature vectors. In addition, they showed that the natural countermeasure of omitting confidence values from model outputs still admits potentially harmful model extraction attacks.
Hu \textit{et al.} \cite{hu2017generating} proposed and generate adversarial malware based on GAN which is able to bypass black-box machine learning-based detection models.

However, several methods have been developed to mitigate both AI-supported and adversarial attacks. However, it is still challenging to implement effective countermeasures against the attacks, as the solutions proposed are still susceptible to other forms of attacks. Moreover, they have not been incorporated into SA systems. %
The Defense Advanced Research Projects Agency (DARPA) has utilized deep learning and neural networks over the past decades to develop machine-speed defensive systems capable of detection, evaluation, and patching vulnerabilities in real-time while probing the attacker's system \cite{DARPA}. DARPA is still working on a program known as Cyber-Hunting at Scale (CHASE) to detect and characterize new attack vectors, collect relevant data, and deploy countermeasures \cite{DARPA_CHASE}. It is a work in progress that seeks to develop an automated tool that will overturn advanced attackers based on both ML and cyberattack modeling tools.




\section{Data Gathering}\label{Data-Gathering}

The capability of a powerful SA system highly related to the quality of data collection. The system needs to collect information about the environment which is mainly from different sources. Then, it can help the system to make decisions based on the collected information and knowledge gained, and consequently response to the threats. Data collection can further improve the quality of knowledge to make better decisions for the future threats.

There are various types of data sources that need to be used for SA, we classify those types of data based on the different criteria such as availability, accessibility, complexity of use, and usability for SA system: {\em dynamic}, {\em one-off}, {\em alert-based}, {\em intelligence sharing}, and {\em raw data}. Each type is detailed as follows:
\begin{itemize}
\item[--] {\em Dynamic}: This type consists of the data which are produced by vulnerability scanning tools (such as NESSUS) or network data gathered through network topology or configuration. The data are usually updated depending on the type of network. However, dynamic data need to be updated frequently once the network components are changed. Changing the network components may change the vulnerabilities that should be captured again. 
\item[--] {\em One-off}: This is usually produced by reports from experts such as incident reports which are mostly static~\cite{fink2013gamification}. 
\item[--] {\em Alert-based}: The examples of Alert-based data are those data produced by Intrusion Detection Systems or other Alert-based systems like Snort, Tripwire, and so forth \cite{sun2017enterprise}.
\item[--] {\em Intelligence sharing}: This can be obtained through communication with other parties or external threats intelligence which have the information such as Open  Indicators of Compromise (OpenIOC) and the Malware Information Sharing Platform (MISP). They usually provide updated information on recent vulnerabilities and malware. 
\item[--] {\em Raw data}: Any other raw data can be categorized in this group such as packet sniffing, system log files, SNMP traps, traffic dumps, OS audit logs, firewall logs, etc \cite{wu2016big}.
\end{itemize}

\begin{figure}[t]
	\centering
	\includegraphics[width=0.97\linewidth]{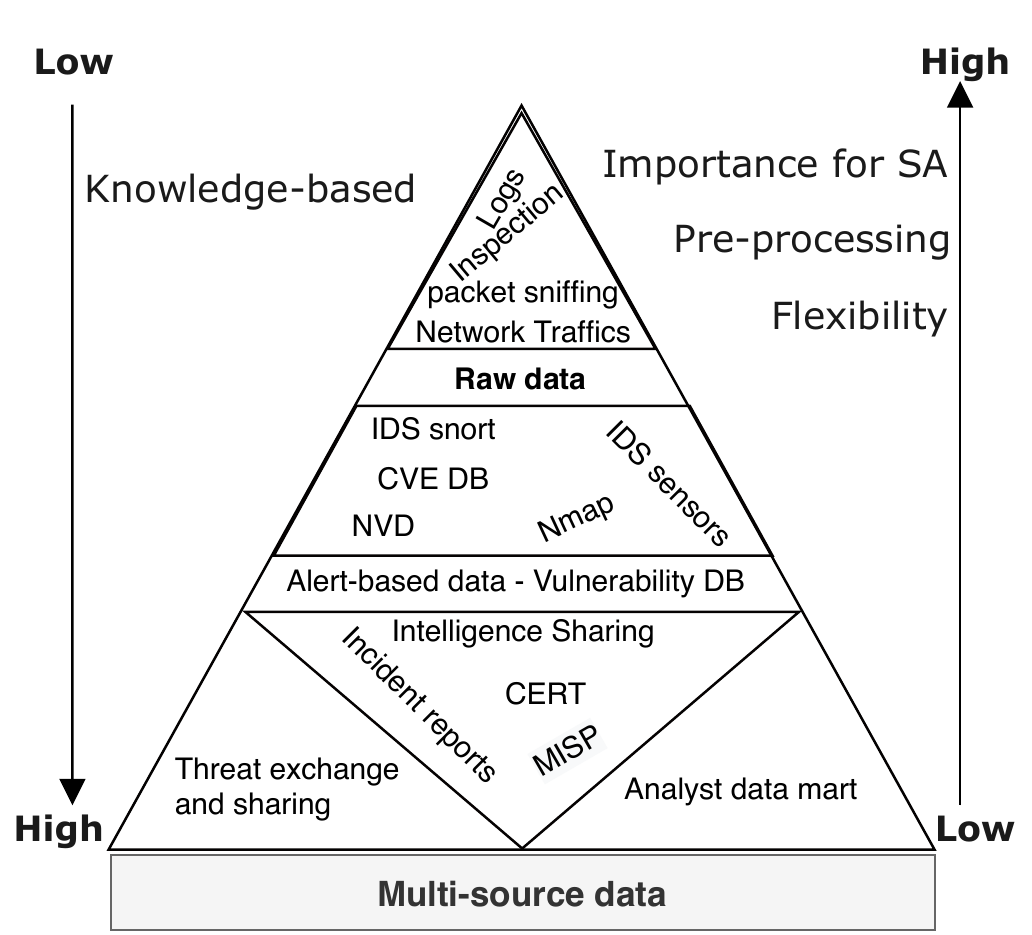}
	\caption{Multi-level data type pyramid model for SA}
	\label{fig:data}
\end{figure}

However, the data types discussed above can be gained through various ways. We further survey how those data types can be collected and used for SA systems which includes the platforms, tools, and resources. Fig.\ref{fig:data} demonstrates the multi-level data type pyramid model for SA.

\subsection{Honeypot} \label{HP}
{Honeypots are systems designed to deceive attackers into believing they are interacting with a real information asset in order to understand attacker behaviour and intentions. The main advantage of honeypots are that they minimise false positives because because a honeypot is not a production asset and so no legitimate user should be accessing it. There are some false positives caused by web crawlers or similar systems such as network measurement tools~\cite{provos2004virtual}}.

The application of Honeypots in designing and creating SA systems have been studies widely~\cite{barford2010employing}. Honypots are used to provide a source of accurate, timely and concise information for SA systems. Honeypots can be used to capture large-scale malicious activity using the traffic inspection, and collect and classify data and fed into intrusion detection system to provide more precise perspective of the current situation of the proposed network. \citet{barford2010employing} proposed a daily network security monitoring system using honeypot to collect the data and further classify and summarize the data to provide ongoing SA. 
\citet{Thonnard2008} leveraged malicious Internet traffic data obtained from a distributed set of honeypot responders (i.e. honeynets) to capture time series of attacks and further clustering of the attack patterns. Similarly, \citet{chawda2014dynamic} proposed a distributed honeypot system aiming to monitor and detect new vulnerabilities.

Moreover, in~\cite{koniaris2014honeypots}, the authors developed a honeypot system to acts as a malware data collector which is able to capture self-propagating malware and monitor their activity. \citet{Sun2019} proposed a novel framework for modelling and clustering attackers' activities using the data collected by world-wide scale honeypot. The collected data from honeypot are fed into further analysis module which uses Bayesian probabilistic graphical model and a graph-based clustering algorithm for classification of attacks and monitoring the attackers' activities.  
Moreover, the application of honeypot in monitoring and detection of botnets has been studies in the literature~\cite{li2008honeynet,fachkha2015darknet}. 
Internet of Things (IoT) development are exploited by the attackers as a viable attack sources for launching various extensive attacks through the botnets. Botnets are appropriate attacking points to launch a wider attack range to any system or network by exploiting the vulnerable IoT-based botnets~\cite{khan2018iot}. For example, a Distributed Denial-of-Service (DDoS) attack can be launched using compromised IoT devices (botnets) to the cloud systems by flooding traffic packets from various sources causing service interruption for users.
Honypots can help to capture malware activities launched from botnets. In~\cite{li2008honeynet}, the authors utilized honeypot to monitor the traffic passing through the honeypot and botnets and extract the malicious botnet activities. \citet{fachkha2015darknet} studied the application of honeypots and darknet to monitor and detect various malicious threats launched from the Internet such as DDoS, Worms, and botnets. 




\subsection{Intelligence Sharing Platforms}
SA systems need to get data related to the threats and malicious activities. Information sharing is a an essential process in detecting security breaches and proactively protecting information systems and infrastructures. SA data (i.e. raw data) can be collected in both lower and higher levels, as data is converted to abstract information. However, the lower levels data can overwhelm decision makers' cognitive capacities. Indeed, relying on low level data solely is obviously insufficient for situation awareness.
For creating automated SA system, data gathering phase should use technical platforms, tools, standards, and secure information exchange protocols to get related data such as higher-level threat intelligence data~\cite{zhang2018network}. However, intelligence sharing tools such as Open Indicators of Compromise (OpenIOC) and The Malware Information Sharing Platform (MISP) or classic data sources like CERTs and Open-Source Intelligence (OSINT) can be used to receive those information~\cite{skopik2016problem}. To create a holistic SA system, a large amount of dynamic data needs to be monitored, refined and processed at real-time.~\citet{yen2010rpd} utilized a hypothesis-driven information gathering (called as ``gathering of evidence") to address the challenge of processing the large amount of dynamic data for building the SA framework.

\subsection{Vulnerability and Network Discovery}
Vulnerability scanning tools can scan network devices and software as well as cloud infrastructure to reveal configuration errors, unpatched and vulnerable devices, and known vulnerabilities~\cite{Alavizadeh2019,alavizadeh2020cyber}. SA can leverage the data collected from vulnerability scanning tools for more analysis. For instance, some vulnerabilities can be patched and some may need additional countermeasures. However, to create a real-time SA system vulnerability scanning tools should perform periodic or continuous scanning process. Moreover, Network scanning can be used to create SA. Network scanning can get network-based information such as devices, platforms, operating systems, and open ports and services, etc. which can be used for SA monitoring system. There are various vulnerability database and sources such as Common Vulnerabilities and Exposures (CVE), National Vulnerability Database (NVD), OSVDB\footnote{https://blog.osvdb.org/}, X-Force\footnote{https://www.ibm.com/security/xforce}, and BugTrack\footnote{http://www.securityfocus.com/}. Vulnerability database mainly includes an unique identifier, description, publishing dat, vulnerability scoring system (such as CVSS\footnote{https://www.first.org/cvss (2018)}), and some security metrics. Nessus, QualysGuard, MaxPatrol, OVAL, and GFI LanGuard, are some examples of tools which have been used in the studies to gain vulnerability information for SA systems~\cite{sun2017enterprise}. 



\begin{table*}[t]
\centering
\footnotesize
\caption{Data types classification based on various factors for SA systems.} 
\vspace{-3mm}
\label{tab:data-type}
\begin{tabular}{@{}|l|l|l|l|l|l|l|@{}}
\hline
\textbf{Data Type}                                                      & \textbf{Example of sources}                                                                                                                                                  & \textbf{Pre-processing} & \textbf{Source} & \textbf{Pros}                                                                                & \textbf{Cons}                                                                                                                            & \textbf{References} \\ \hline
\textbf{Dynamic}                                                         & \begin{tabular}[c]{@{}l@{}}-- vulnerability scan data\\ -- network topology\end{tabular}                                                                           & Moderate & {I} &                 \begin{tabular}[c]{@{}l@{}}+ Available\\ + Include metrics\end{tabular}                                                                                   & - Need Frequent update                                                                                                                   &    \cite{albanese2017computer}                 \\ \hline
\textbf{One-off data}                                                    & -- Incident reports                                                                                                                                                & High   & {I}--{E}--{H} &                + More precise                                                                               & \begin{tabular}[c]{@{}l@{}}- Less availability\\ - Need experts\end{tabular}                                                             & \cite{fink2013gamification,kanstren2016study}                    \\ \hline
\textbf{Alert-based}                                                     & \begin{tabular}[c]{@{}l@{}}-- IDS\\ -- alerts (e.g., Snort alerts,\\ tripwire alerts)\end{tabular}                                                                 & Low & {I}--{H}                    & \begin{tabular}[c]{@{}l@{}}+ No inspection required\\ + Real-time (alert based)\end{tabular} & \begin{tabular}[c]{@{}l@{}}- Inability to detect\\   unknown attacks\end{tabular}                                                        &   \cite{sun2017enterprise}                  \\ \hline
\textbf{\begin{tabular}[c]{@{}l@{}}Intelligence \\ sharing\end{tabular}} & \begin{tabular}[c]{@{}l@{}}-- CERT\\ -- MISP\\ -- OSINT\end{tabular}                                                                                               & Low   & {E}                & \begin{tabular}[c]{@{}l@{}} + More precise \\ + Updated Information\end{tabular}                                                                                   & \begin{tabular}[c]{@{}l@{}}- Need external sources, \\ sharing, and trust, \\ so less available and slow\end{tabular}                    &  \cite{kanstren2016study,zhang2018network}                   \\ \hline
\textbf{Raw}                                                             & \begin{tabular}[c]{@{}l@{}}-- Packet sniffing\\ -- system log file\\ -- SNMP traps\\ -- other like traffic dumps, \\     OS audit logs, firewall logs\end{tabular} & High  & {I}--{H} &            + More available                                                                             & \begin{tabular}[c]{@{}l@{}}- Changes Frequently\\ - Complexity i.e. parsing \\    and analysis\\ - Needs frequent inspection\end{tabular} &   \cite{wu2016big}                  \\ \hline
\end{tabular}
\vspace{-0.5mm}
\begin{flushleft}\textbf{~~Notations}: Internal Sources (\textbf{I}) - External Sources (\textbf{E}) - Honeypot (\textbf{H})\end{flushleft}
\end{table*}

However, network and system information such as topology, configuration, components, etc. are important in monitoring the network's activities based on the available components and connectivity. Network flows can be collected and be correlated with security events. They may also be useful for attack representation techniques such as attack trees and attack graphs~\cite{kotenko2018ai}. Various network discovery and mapping tools such as Lumeta IPsonar, SteelCentral NetCollector (formerly OPNET NetMapper), Nmap or JANASSURE are useful for SA and have been used in studies~\cite{sun2017enterprise}. They screen the incoming and outgoing traffic in the network and monitor crucial files on the host operating system.

\subsection{Network Traffic Inspection}
Network traffic inspection have been used in many studies to collect the required data for creation of SA systems. \citet{Vinayakumar2018} conducted DNS data collection in a passive manner by using using promiscuous mode and reading the mirrored traffic on DNS communication between both DNS clients and servers. The data includes DNS queries and the DNS answer regarding each query made by the DNS communication between the client and DNS server. They used the collected data such as malware propagation and activities, the prefix announcements, route announcements and updates information to identify the malicious activities. Moreover, Antivirus software (AVS) which are traditional countermeasure can be used to collect data for SA creation. Many AVS can produce log data about detected malware and are able to generate log data about network traffic which can be utilized by SA. \citet{wu2016big} conducted a data extraction methods based on the factors and rules which are related to the design of the proposed SA system for smart grid. The rules of extraction were defined based on the system's requirements. Situational factors store the security related information and heterogeneous schemes according to the specified format. They collected the semi-structured and unstructured data required for SA using the basic situational factor collection such as network flow, access control operations, and device states. They performed data collocation based on network traffic inspection by inspecting Network Management Protocol (SNMP) data flows which manages TCP/IP network communications. Some tools such as Snort, TCPdump, Bro, Ntop, and WireShark are some examples of network traffic inspection tools which have been used in the studies to gain network traffic-based information such as log and traffic flows for SA systems~\cite{sun2017enterprise}. 



\subsection{Intrusion Detection System (IDS) \& Firewall Data}
{IDS plays an important role on monitoring the individual devices and the system's network traffic. Various monitoring techniques can be leveraged by IDS such as log events monitoring and analyzing, and signature and anomaly based systems. For instance, packets IP addresses, action sequences (remote user logins and manipulating the critical files) can be the examples of signature-based monitoring. However, reference usage profile is categorized as the anomaly-based monitoring. For instance, when a normal user such as the office secretary uses the tools which are only used by admin stating team, it can be considered as abnormal usage behaviour and might be detected. Moreover, IDSs should be are able to comprehensively monitor traffic anomalies to gain a clear situation awareness based on the sources, destinations, and the amount of traffics. Moreover, the other aspects such as security policy violations and system integrity can also be detected by IDS.} Some IDS such as TripWire, OSSEC HIDS, and Snort have been used in the studies for gain abnormal events on both network and individual hosts.





\subsection{Data Gathering Limitation} As the most essential part of SA system design, data gathering process is still crude and need further consideration. 
More cost-efficient data gathering and pre-processing need to be investigated. The limitations of data collection for SA are listed as follows~\cite{kotenko2018ai}, and also pros and cons of each are illustrated in Table~\ref{tab:data-type}:

\begin{itemize}
\item[--] Different data types need various sources which increase the complicity of data collection and real-time monitoring capabilities.

\item[--] Although honeypots and honeynets are crucial sources of gathering information about real threat scenario, most of the gathered data using Honeypots are unstructured and unorganized. It's crucial to automate the data organizing and condensing of honeypots to be useful for further SA evaluation and real-time analysis.

\item[--] Various sources provide different data representation and formats which increase the pre-processing burden. However, devising a standard and unique dataset which can incorporate data types from different sources into a standardized form is still missing in the literature.

\item[--] Continuous growth of data volumes collected and stored for further analysis (i.e. using honeypots) may cause problems by overwhelming the system. However, a system should be designed to remove the unnecessary data and preserved some portion which may use for further process (such as learning purposes). Thus, a systematic data gathering is needed to extract only useful information from large traffic data sets.
\end{itemize}

\section{Analysis and Techniques} \label{Analysis}
\begin{figure*}[t]
	\centering
	\includegraphics[width=1\linewidth]{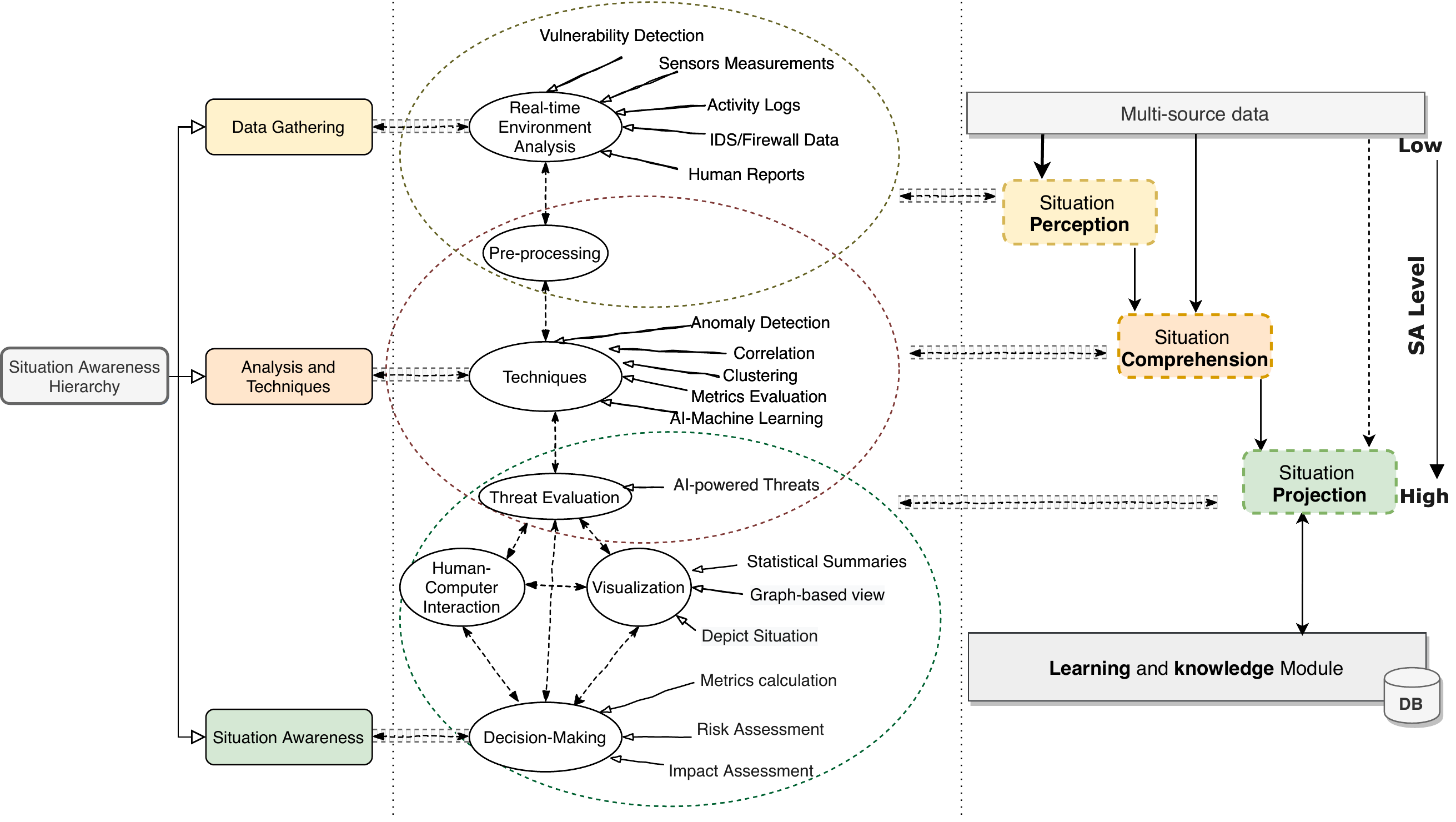}
	\caption{Situation Awareness hierarchically structure and the related framework showing the level of situation awareness.}
	\label{fig:taxonomy}
\end{figure*}

This section provides a comprehensive review and classification of the existing methods and techniques used to analyze cyber SA in various systems and contexts. Gaining a high level of situation awareness depends on three main hierarchical steps each of which provide a level of situation understanding from low to high. SA phases can be classified into three main categories including data gathering, analysis and techniques, and situation awareness demonstrated in Fig.~\ref{fig:taxonomy}.


\citet{he2017survey} defined the three main layers of awareness for a comprehensive SA monitoring system as follows. The main task of the first layer is to monitor and analyze abnormal traffic of the network through network traffic monitoring and analysis tools and called it as data extraction layer. For this purpose, the alarm database can be used to extracted sample data and then the sample data can be processed. The second layer is the key layer of the system which is used to realize and evaluate the network security situation and determine whether if the network is attacked or not. Consequently, it processes and evaluate attacks based on the current system situation and based the evaluation model of the specific situation, it generates the security situation based on the corresponding graph to reflect the current status and security posture of the network. Finally, the third layer of the system can evaluate and predict the network security situation based on the second layer's output. It predicts the network security situation, by obtaining the current network security situation and other network security data. This stage can help the security experts to be aware of the network security situation with more high-level information, and also help them to provide the basis for making reasonable decision.

\subsection{{Data Pre-processing}}
{All collected data, especially raw and unformulated data, need to be parsed, cleansed, normalized before feeding to the next step for further analysis. Indeed, the data pre-processing consists of various steps such as data cleansing including duplicate elimination, data normalization and collation. For instance, data cleansing may include data calibration and filtering process for the raw data collected from security sensors (IDS, network and system log records, firewall, SIEM, and NetFlow and so on}. Data processing for SA system design and implementation has been discussed in various studies~\cite{tianfield2016cyber, zhong2017studying}. \citet{zhong2017studying} conducted an extensive study on the data triage operations in situational awareness analysis. Their main area of focus are in cyber defense analysts based on the data triage and network monitoring data, their proposed framework helps to automate the retrieval of data triage process for analysts.

%
\vspace{1mm}
\noindent\textbf{Data Refining and Fusion:}
Most of collected input data, especially for the raw data collected by IDS or network traffic inspection, log files, etc, as shown in top of pyramid in Fig. \ref{fig:data}, need higher pre-processing burden. It mainly includes parsing and refining of raw data before feeding the refined data to analysis modules. For instance, \citet{shen2007markov} defined three levels for collected data: (i) Object refinement which mainly are pre-process and refining raw data collected through various sources such as IDS, system and web log files, as a level-1 data fusion, (ii) Situation refinement or level-2 fusion which deals with pre-processed data resulting from level 1, and (iii) Threat refinement or level-3 data fusion which consists of high level data resulting from level-2 data fusion. This hierarchical perspective of data fusion is a part of data processing to determine final situation awareness and designed impact assessment system. Similarly,~\cite{yuan2015data} proposed a data fusion model for Resilient Control System which aggregates data form various IDS sources and pre-process them using two levels as object refinement which uses a combination of IDS to identify an individual attack. Then, the results from the object refinement are fed into the situation refinement which can find simple attack characterization by fusing the defensive posture. Later on, the pre-processed and fused data from those two levels should be fed into higher analysis based levels such as threat refinement. In~\cite{tao2016big}, the authors proposed a suitable data fusion algorithm for SA aiming {to reduce network traffic redundancy in big data through feature extraction, classification, and integration. Feature extraction is The key component of the data fusion algorithm which helps to address dimension complexity. The features can show the original big data through analyzing the internal characteristics.} 



However, cybersecurity problems are, often, compresses a very large set of features. Utilising all the features has two main drawbacks, firstly, more examples will be needed to build/train a model, and secondly, not all features are relevant or needed. Hence, performing feature selection and feature construction can largely help to mitigate this problem and reduce the dimensionality of the feature space. Furthermore, the vast majority of those features are handcrafted by domain experts. Seeking an automated approach to extract features from the raw date or building a model that is capable to operate directly on the raw data can reduce the effort and costs associated with manually designing and extracting those features.

\subsection{Using Artificial Intelligence} \label{subsec:_tech_suffling} \label{AI-ML}

\subsubsection{Machine Learning}
{
	In computer science, machine learning (ML) is a subfield of the wider artificial intelligence field (AI) \cite{Segaran:2007} and one of the most rapidly evolved fields \cite{Alpaydin:2014}. ML comprises a set of algorithms that aim at automatically, i.e., without being explicitly programmed, extracting useful patterns or knowledge from data to solve a problem \cite{Alpaydin:2020}.
	Generally, ML methods can be categorised into four approaches: (1) supervised learning, (2) unsupervised learning, (3) semi-supervised learning, and (4) reinforcement learning \cite{Weiss:2016}.}
{
\begin{itemize}
\item[--] {\em Supervised Learning}: The methods of this approach relies on labelled examples, i.e., the correct output (also known as ground truth) is provided, and are guided by those labels during the training process to learn a model. Hence, such methods aim at generating a generalised mapping between the inputs and the corresponding label \cite{Bishop:2006}. The two typical tasks of this approach are classification and regression.
\item[--] {\em Unsupervised Learning}: The methods of this approach concern with unlabelled example inputs and therefore, the main aim is to utilise predetermined criteria to group those examples into different groups. A typical task of this approach is clustering.
\item[--] {\em Semi-supervised Learning}: The methods of this approach combine the both of the previous two approaches. Such methods are needed where the problem at hand comprises a mixture of labelled and unlabelled data. Transductive support vector machine is a typical example of this approach \cite{Joachims:1999}.
\item[--] {\em Reinforcement Learning}: The methods of this approach aim at developing an agent that can automatically explore an environment and takes an appropriate action. The learning process in of such methods rely on maximising a cumulative reward and minimising the penalty.
\end{itemize}
}
%
%
%
%
	
\citet{adenusi2017development} developed a threat detection model aiming to gain cyberspace condition. They proposed a SA model using Artificial Intelligence (AI) technique. They used Artificial Neural Networks (ANN), also called (NN), to create perception sub-model. Moreover, they leveraged Rule-Based Reasoning (RBR) techniques to model SA comprehension and projection phases. \citet{champaneria2014survey} proposed a model for detection of novel and unknown attacks. They developed an intrusion detection model using hybrid Artificial Neural Network (ANN) approach. They showed that the proposed hybrid ANN model outperforms the other methods in terms of attack classification, training time, and detection rate. 

{SA prediction is one of the crucial purpose of SA system. Various methods of network security situation prediction have been proposed based on different ANN models such as Back Propagation (BP) Neural Network (BP-NN)~\cite{zhang2012network,tang2011security}, Radial Basis Function (RBF) Neural Networ~\cite{zhang2012network}, Elman Neural Network (ENN)~\cite{zhang2014predicting}, and etc~\cite{meng2011network}. For instance, \citet{tang2011security} proposed a network security SA prediction method based on dynamic BP neural network using covariance. They improved their method by including self-learning dynamic adjustment of the parameters' weight.}

{\citet{zhang2012network} proposed a prediction-based SA based on two neural network models: BP-NN and RBF-NN. They also compared those two methods and showed that BP-NN model is more effective than RBF-NN model to predict network SA. \citet{zheng2012strategy} proposed a self-adaptive and real-time SA strategy named Network Security Situation Autonomic Awareness (NSSAA). They adopted a BP-NN model to realize self-learning adjustment of input data.} 

{A hybrid ANN models for SA is proposed by \cite{zhang2014predicting}. The authors combined three NN models which are BP-NN, and RF-NN together for predicting SA for a computer network. They showed that the combination of NN models can yield better and more efficient SA prediction.}

{However, in most of proposed ANN models, the authors performed error analysis and predict error values, the error values later used for training purposes of prediction models. Thus, the improvement on the models could only achieved by previous prediction errors fed into the prediction model as a training sample. Moreover, the success of the ANN models highly depends on training sample, algorithms, and the quality of training~\cite{leau2015network}.}

{The application of supervised learning for SA prediction has been studied in~\cite{chen2013network}. \citet{chen2013network} used echo state networks (ESNs) as a supervised learning method with small world property to propose a network security prediction method by training the historical attack records. Moreover, in \cite{hu2019improved}, the authors proposed a method for SA prediction model based on a support-vector machines (SVM which can learn based on training of large volume of input data using KDD dataset. They showed that the model effectively reduces the SVM training time while it enhances the accuracy of SA prediction.} 

\citet{vinayakumar2019deep} used different deep learning architectures to be able to detect spam and phishing attacks using Uniform Resource Locator (URL) and email data sources due to the importance of Email and URL resources which are used by the attackers to spread malware. They used various datasets to conduct their experiments using deep learning architectures. They used classical machine learning algorithms for comparative study and collected required data using public and private data sources. \citet{Vinayakumar2018} proposed a scalable framework for situational awareness for networks which is able to perform web scale analysis in near real-time manner and detect threats and emit early warning signals to avoid malware propagation and large scale attacks. They employed deep learning approach to correlate malicious activities obtained from the DNS protocol usage. \citet{Dietterich2010} used machine learning methods to capture the behavior of ordinary desktop computer users.

{However, the SVM-based approaches is effective for SA modeling and analysis including monitoring and prediction of SA. However, most of these models suffer from the long training time mainly for SA prediction model which is the main drawback of these types of ML-based approaches~\cite{hu2019network}.}


%
{\subsubsection{Evolutionary Computation}
	Evolutionary computation (EC) is concerned with biologically inspired algorithms \cite{Back:1997}. EC algorithms can be widely categorised into two groups \cite{Back:1997}: Evolutionary Algorithms (EAs) \cite{Jong:2006}; and Swarm Intelligence (SI) \cite{Eberhart:1995,zhao2018study}. EC algorithms are considered to be global searchers as they rely on a population of candidate solutions unlike other ML algorithms that search the space using a searcher.}
%

\citet{liang2007quantification} proposed a SA system based on incorporation of evolutionary algorithms and neural network models. They used the evolutionary algorithm to optimize the parameters of neural network model and finally quantify the network SA. However, their model could only analyze limited situational factors which are mainly based on SA perception. Later on, \citet{lin2008pso} proposed a SA model by incorporating BP Neural Network and Particle Swarm Optimization (PSO) to predict future situation and projection. The incorporation of PSO into the model can provide global optimization solution. However, it lacks generalization for new samples and only relied on trained samples. Similarly,~\citet{meng2011network} proposed a SA network security prediction method based on combining RBF-NN with hybrid hierarchy genetic algorithm. \citet{li2013situation} proposed a SA extraction method based on the improved particle swarm optimization (IPSO) and logistic regression algorithm (Logistic Regression LR) which can find global optimization and improve the learning speed and accuracy because of the intrinsic parallelism and optimization capabilities of IPSO. Moreover, \citet{zhao2018study} proposed a SA system based on adopting PSO into wavelet-NN model for monitoring large scale data environment and find the optimal solution. Finally, they showed the effectiveness of combining ANN models with PSO methods to achieve faster and more accurate network situation awareness.

Moreover, Combining ANN methods with EC-based methods such as PSO can enhance the SA system in terms of multiple factors such as NN ability to learning, speed of learning, accuracy, and effective solution, while it covers the limitations of traditional NN algorithms such as network training errors and low search success rates~\cite{zhao2018study}. 

\subsubsection{Limitation of AI-based Techniques}
Although ML-based and EC-based approaches are useful in terms of self-learning, automation capabilities of SA , and the ability of combining with other methods, there are some main limitations summarized as follows:
\begin{itemize}
\item[--] Dependency to updated dataset which includes data for novel and unknown threats for training purposes. However, the role of honeypots for capturing real and novel threats is inevitable to address this shortcoming~\cite{brynielsson2016cyber}.
\item[--] Lack of adequate training samples or appropriate training models may cause undesirable results~\cite{leau2015network}. Trained samples should be productively used as the input of model to be able to capture or predict the incoming security situation. Evolutionary Computation methods mostly need a large amount of prior knowledge to extract the situational elements which might be difficult to obtain.
\end{itemize}



\subsection{Game Theory}
Game theory (GT) is related to the mathematical models that can be used in cyber SA to study the game behaviors between attackers and defenders~\cite{blasch2015review}. In here, we highlight the GT-based approaches regarding either designing or applications of game theory by reviewing the literature to provide the current state-of-the-art of GT in SA systems and it's limitations.

Indeed, the main goal of the game-theoretic SA approaches is to predict the adversary behavior against defender. This prediction can be used to provide an advantage to the defender~\cite{albanese2017computer}. Through game-theoretic analysis, the defender can theoretically prove the attacker's best strategy; consequently, the best defensive strategy can be used. Several game-based security awareness methods have been proposed~\cite{zhang2018network,zhang2011network,li2020research}.

Various studies utilized traditional GT-based approach for SA and monitoring systems. \citet{shen2007markov} proposed another Markov Stochastic Game Model for designing SA system which is able to estimate and detect cyber attack pattern based on the collected data. They showed that their proposed GT-based model can enhance the understanding of the network situation and help with proper defense.~\citet{wang2008stochastic} proposed a stochastic game theory model for quantification of network situation awareness. They used the network offense and defense game based on the network service states to realize the payoff of both sides and quantify the situation. However, applying the game theory based on the network state spaces has scalability issue as the network state spaces could be extremely large, especially, in dynamic networks and the solving the state combination problem would be time-consuming. Later on, a stochastic game theory for SA was proposed by~\cite{ying2010dynamic} which could capture larger network states in a dynamic network environment. However, it still suffers from scalability problem in the larger network states to solve state combination problem. 

The applications of GT-based evaluation which can be utilized for SA models have been studied in the literature~\cite{charles2010belief,alavizadeh2021evaluating}. In~\cite{white2013game}, the authors implemented a GT-based model in order to analyze several attack scenarios on Online Social Network (OSN) to obtain clear perspective of system's vulnerabilities against various attacks, and consequently, provide protection mechanism to avoid the attacks. They applied Markov decision process (MDP) in their GT model to secure information sharing in online social networks. 

\citet{zhang2011network} proposed an approach to improve SA based on the Markov Game Model (MGM) by gaining the data regarding threats, assets, and vulnerabilities and evaluating them in real-time. In their model, users, network administrators, and attackers establishes three players for MGM. They showed that the evaluation result is efficient and precise. Moreover, \citet{zhang2018network} leveraged a game-theoretic approach to defend threats in cloud environment using threat Intelligence. They used the Nash equilibrium together with fuzzy optimization method to predict the attack behavior. \citet{ying2010dynamic} proposed a game-theoretic based dynamic SA system by modeling both attacker and defenders and game player using stochastic game model of the network. Then, they quantified the network SA by incorporating the game mathematical formula, attacker and defender costs. They used Nash equilibrium to find a balance between attacker's and defender's benefits.

However, GT-based approaches have some limitations such as lack of full rationality of the involved players which are attackers and defenders and incompleteness of information~\cite{chung2016game}. When it comes to advanced adversaries, this problem even get worse as the intelligent attacker may learn about the defender and the environment. To this end, GT-based approaches need to be incorporated by learning approaches and knowledge awareness to be able to increase monitoring and defensive capabilities against advanced adversaries.

\subsection{Hybrid Approaches}
Various studies have applied multiple approached for designing SA system such as combining learning-based techniques with GT~\cite{wu2016big,xing2016network,watkins1992q,xing2016network}. 
\citet{wu2016big} proposed a security situational awareness using Fuzzy cluster based analytical method, game theory, and reinforcement learning mechanism. They analyzed the big data in the smart grid aiming to gain the security situational awareness in the smart grid environment.
\citet{xing2016network} proposed a method to determine the situation of the system based on Fuzzy Dynamic Bayesian network. Their simulation results were compared with static  Bayesian  network model. They showed that the proposed method can better reflect the dynamic changes on the network.
In~\cite{adenusi2017development}, the authors used the fuzzy equivalent relations in order to perform cluster analysis together with the association analysis the situational factors in big data. 

Moreover, \citet{chung2016game} proposed a hybrid approach incorporating GT and a model-free reinforcement learning (called Q-learning~\cite{watkins1992q}) to enhance the monitoring and defensive capabilities of the designed system and support the problem created by GT-based approaches which are mainly lack of information about the ability or intent of attackers.

However, there still lack of appropriate training dataset for learning-based approaches which have been various application when they are combined with other approaches such as GT and Fuzzy models. Thus, it is necessary to provide further study on improving learning-based and hybrid approaches using real data, and to test the approach with real-time traffic logs, data, and specific detection using a real adversary data collection testbed such as Honeypots.

However, the SA monitoring quality on effectively detecting and analysis of system's environment is still limited due to various reasons including, (i) inaccurate and incomplete data gathering and pre-processing, vulnerability analysis, intrusion detection, (ii) the ability to quickly and automatically adapt to the evolving and dynamic nature of networks, (iii) the capability to deal with intelligent attackers and AI-powered threats (such as sophisticated and complex attacks), (iv) limited capability to deal with uncertainty, and (v) limited capability for large-scale real-time data collection system. 

{\subsection{Triage Analysis}
Much like the triage in medical use, the triage used in the cybersecurity refers to a set of automated techniques with the capability to quickly assess a security incident to determine if the security incident requires further investigation. The term triage has especially become popular in malware detection as an efficient mechanism which can analyze and identify specific malware that require urgent attention from the massive amount of malware for many organizations with limited resources. By and large, most triage techniques that are currently on offer goes through a number of phases, including (1) feature extraction, (2) similarity measure, and (3) clustering. Most often, features are extracted from malware. The features are often clustered according to the result obtained by a similarity metric. The features from the different cluster are compared with known signatures. If no match is found with the known signatures, they can be often classed as potential zero-day attacks. The zero-day attacks are typically highly ranked for further investigation. 
}

{
The current triage techniques can be broadly classified into two groups, the ones that deal with malware features extracted as categorical data while the other deals with malware binary files. BitShred~\cite{jang2011bitshred} and VILO~\cite{lakhotia2013vilo} extract features based on N-grams that have been used in text analysis. BitShred further use a feature hashing~\cite{shi2009hash} on the extracted feature to allow for dramatic dimensionality reduction to compress large feature space down to a smaller feature so the hashed presentation of features take less space in memory and more effective for cache. The hash features are compared using the Jaccard similarity metric in BitShred while VILO utilizes a weighting scheme~\cite{jones1972statistical} by calculating how frequently a word (i.e., feature) appears in the feature vector and comparing the similarity in the weights. BitShred uses co-clustering method to correlate both the hashed features and malware samples which are claimed to discover more substantial, non-trivial structural relationship among malware samples. VILO uses the nearest-neighbor algorithm based on the weighted similarity scores to form clusters. Instead of N-gram, MAST~\cite{chakradeo2013mast} extracts features based on the qualitative data that represents each mobile app (e.g., permissions, intent filters, the presence of native code, etc.) -- this is named as questionnaire. The similarity calculation among the collected questionnaire is done using a statistical method called Multiple Correspondence Analysis (MCA)~\cite{abdi2007multiple} that measures the correlation between multiple qualitative data followed by grouping related mobile apps together (i.e., clustering) so that uniqueness within the group is more specific. In contrast, SigMal~\cite{kirat2013sigmal} uses the signal processing-based feature extraction where the executable binary content as a one-dimensional signal that is represented as a vector of bytes. The vector of bytes is converted as filtered feature vectors~\cite{nataraj2011malware}. The authors claim that the use of the filtered feature vectors based on the executable binary content is better equipped to preserve the features of the original malware sample even though the malware is disguised by polymorphic engines or general packers (e.g., encryption or compression techniques). Euclidean distance metrics is used between feature vectors to find the nearest-neighbor sample in the learning dataset. 
}

{
The main point of research in triage techniques is either to improve detection accuracy which is decided by the algorithms utilized for similarity metrics and clustering or to speed up computation to filter through as many samples as possible. For example, BitShred showed that the proposed method speeds up typical malware triage tasks by up to 2,365x and uses up to 82x less memory on a single CPU thus more suited for large-scale malware triage and similarity detection. The results from VILO presents that there was in between 0.14\% and 5.42\% fewer mis-classification compare to similar methods. MAST was able to detect 95\% of malware from the 36,710 mobile apps as test samples. SigMal could classify 50\% of the incoming sample with above 99\% precision and showed that it could have detected, on average, 70 malware samples per day before any antivirus software detected them.
}

\subsection{Anomaly Detection}\label{AD}
Anomaly detection can be incorporated with SA to detect abnormal behaviour of a system components such as users, traffic, access, etc. and provide the system with more clear idea about the current situation of the system based on normal and abnormal activities.
Actually, abnormal behaviors are the activities inside a system which are opposite of the normal or logistic behaviors. Thus, the core of anomaly detection is established based on the monitoring of normal system operation to find out any deviation from the normal model~\cite{li2019analysis}.

\citet{friedberg2015cyber} distinguished three kinds of anomalies including (i) point anomalies which involves with a single event that can be considered anomalous given the notion of normality, (ii) contextual anomalies which refers to when an event can be considered as anomalous behavior for a given context. Thus, the anomaly can be inferred using the the events' behavioral attributes in its context, (iii) collective anomalies which indicates a series of events which are considered as anomalous activities. However, in order to precisely assess the current situation of the system, the SA system should be able to detect all those three discussed abnormality behaviors using anomaly detection methods embedded in the SA system. The application of anomaly detection in SA has been studies in the literature~\cite{harrison2012situ,li2019analysis,mcelwee2017active}. For instance, \citet{harrison2012situ} proposed an anomaly detection method to detect low probability events for SA.

\subsubsection{ML-based Approaches}
ML-based approaches have been significantly used for gaining situation awareness through anomaly detection~\cite{mcelwee2017active, salama2011hybrid}. Various ML techniques have been used such as the symbolist approaches using random forests~\cite{mcelwee2017active} and decision trees, or connectionist approaches leveraging neural networks (NN)~\cite{mcelwee2017deep}, or evolutionary approaches which mainly mimic genetics or the immune system~\cite{dasgupta2002immunity}. Moreover, other techniques such as Bayesian methods~\cite{valdes2000adaptive} and analogistic approaches using support vector machines (SVM)~\cite{mukkamala2002intrusion} have also used in the literature. However, learning based approaches need to use existing dataset for training and testing purposes. We further explain two popular dataset used for this purpose.  

\vspace{1mm}
\noindent\textbf{Datasets:}
Most of the learning based anomaly detection techniques use two datasets such as KDD-Cup 1999 dataset and NSL-KDD dataset which are more popularly employed as training and testing datasets.

\begin{itemize}
\item[--] {\em KDDCup 1999~\cite{tavallaee2009detailed}}: {This dataset is a popular dataset which has been widely utilized for intrusion detection and anomaly detection methods. The training dataset includes around 5,000,000 single connection vectors which contain 41 labeled features as two types of normal or an attack. The features labeling as the Attacks are based on four categories such as DoS, User to Root Attach(U2R), Probing, Remote to Local Attack (R2L).}

\item[--] {\em NSL-KDD~\cite{kayacik2005selecting}}: NSL-KDD dataset is another dataset which has been used in ML-based methods. This dataset addresses the shortcomings of the KDDCup 1999 dataset. The KDDCup 1999 dataset includes a large amount of redundant or duplicated data records which are around 75\% and 78\% in both testing and training dataset respectively. This redundancy could make the learning algorithm bias and cause wrong results. To address this problem, NSL-KDD is adopted as the new version of KDDCup 1999 dataset and widely adopted for anomaly detection.
\end{itemize}

\vspace{1mm}
\noindent\textbf{Feature Manipulation:}
Feature manipulation is an important data pre-processing step for anomaly detection, especially for classifying high-dimensional data. It mainly refers to the process of transforming the input space of a machine learning task aiming to enhance quality and performance of learning-based techniques such as Machine Learning (ML). Feature manipulation concerns with feature selection, feature construction, and feature extraction. Feature selection aims at selecting a subset of the original features by removing irrelevant, redundant, and noisy features \cite{Xue:2016}. Feature construction aims at generating a new feature or set of features by considering various combinations of the original features \cite{Neshatian:2012}. 

\citet{fiore2013network} proposed a network anomaly detection in a semi-supervised fashion based on the Discriminative Restricted Boltzmann Machine (DRBM)~\cite{larochelle2008classification}. They used DRBM to capture the main aspects of the normal traffic class and further perform accurate classification. Another learning-based technique was developed by \citet{salama2011hybrid}. They conducted research for the anomaly intrusion detection scheme using deep learning methods called Deep Belief Network (DBN)~\cite{deng2014deep}. They leveraged SVM classifier together with DBN used for feature reduction and called a hybrid scheme of DBN and SVM. Their hybrid methodology includes three main phases which are pre-processing, DBN feature reduction, and classification.
\citet{iglesias2015analysis} proposed an anomaly detection for network traffic based on the feature selection approaches. They utilized a multi-stage feature selection method using filters and step-wise regression wrappers. Then, more advanced anomaly detection model based on deep learning was proposed by \citet{javaid2016deep}. They developed an anomaly detection-based system (ADNIDS) to detect unknown future attacks using deep learning approach. They introduced two main steps for feature extraction to collect unbalanced network traffic data and supervised classification to use the extracted features to label traffic dataset. They used NSL-KDD dataset for training data and evaluated the performance of the approach using some metrics such as accuracy, precision, recall, and f-measure values. 

Further in \cite{kim2016long}, the authors proposed a hybrid anomaly detection model based on incorporating Long Short Term Memory (LSTM) and Recurrent Neural Network (RNN). They trained their proposed deep learning model using KDD Cup 1999 dataset and showed the effectiveness of their approach to detect the attacks in comparison with other learning based anomaly detection techniques. Similarly, \citet{tang2016deep} utilized Deep Neural Network (DNN) model for anomaly detection in Software Defined Networking (SDN) environment. They used NSL-KDD Dataset for training their deep learning model and showed the effectiveness of DNN to monitor and detect various attacks in the SDN environment.


\subsubsection{White-list and Black-list Analysis:}
Blacklisting is a classical approach helping a monitoring system to detect malicious activities by maintaining a list of known blacklisted threats or activities. In contrast, white-list technique is used to gather and classify the information of reliable sources for legitimate uses~\cite{vinayakumar2019deep}. The application of white-list or black-list techniques have been studied in the literature~\cite{ezick2019combining,bradshaw2012sol,nakakoji2016proposal}.
In~\cite{nakakoji2016proposal}, the authors proposed an authentication-based approach which classifies the access to the Uniform Resource Locators (URLs) based on three defined lists: white-list, black-list, or gray-list. They showed that the proposed system can improve the accuracy of the suspicious the gray-list, white-list, and black-list, and further reduce the authentication frequency for the user accessing the URLs. 

However, there still a limitation for the accurate classification of those lists. For instance, a legitimate URL may be misclassified as a blacklist or vice versa. Moreover, most of the white-list and black-list approaches need frequent updating as the thousands of emerging threats evolve every day, and updating this list would be challenging~\cite{jain2016novel}. 
\begin{itemize}
\item[--] {\em DNS-based Black-list:} Domain Name System (DNS) can contains resource records for the identification of hosts presented in the black-list and uses the DNS protocol.
\item[--] {\em Botnet Detection}: \citet{prieto2011botnet} proposed a Botnet detection system called as Botnet Detection System (BDS) which includes the network tools such as wget, Net-Whois, dig, and perl script to analyze the DNS traffic. They used a test-bed system which was infected with Zeus, Conficker and Kraken botnet to obtain the Black-list data.
\item[--] {\em Firewall and Access-list}: In~\cite{fitzgerald2013mason}, the authors defined a black-list for smartphones to avoid sending and receiving traffics to a known malicious host. They also defined a white-list for the legitimate apps allowed to connect to the network connection. 
\end{itemize}

\subsubsection{Endpoint Protection:} 
Endpoint protection is used to describe a set of security solutions designed to secure endpoints or entry points of end-user devices (e.g., laptops, tablets, smart phones, and other wireless devices) that are used to connect to the organization networks. In recent years, the organization has increasingly contended with not only growing number of endpoints but also a rise in the number of types of endpoints (e.g., IoT). Compounded by remote work and BYOD policies, these factors have created more wide attack paths making endpoint security more difficult and traditional firewall and antivirus-based approaches increasingly insufficient. 

{
Endpoint protection can be characterized to attempt (1) securing the entry points of end-user devices, (2) protecting endpoints on a network or in the cloud from threats. For the former, a number of user authentication mechanisms have been proposed to ensure only legitimate end-user devices are connected safely to organization networks. Mutual authentication\footnote{https://tools.ietf.org/html/rfc5246} supports a mechanism where both entities (i.e., a client and a server) authenticate each other, either based a certificate exchange or username/password verification. Open Authorization (OAuth)\footnote{https://oauth.net/2/} has become one of the most popular and widely used authentication mechanisms on the Internet as it allows a federated user authentication where a client can use a single authentication token to assess a several organizations across trust boundaries. With the growing concern on IoT devices increasingly connected to organization network (via Fog or Edge computing), many monitoring systems incorporate capability to authenticate IoT devices. \citet{almadhoun2018user} proposed a decentralized and scalable authentication approach that utilizes blockchain-enabled connectivity to Ethereum smart contracts where access tokens to communicate to the organization network server are issued by the smart contracts with no intermediary or trusted third party by effectively removing the overhead and expense associated with the third party solution. Advanced biometric-based approaches to bind an end user with his/her registered mobile devices (e.g., Apple iPhones) to generate a device unique “fingerprint” and use the unclonable fingerprint to authenticate with the server has been proposed~\cite{guo2017voice,zheng2019udhashing}. For the latter, NICE~\cite{kienzle2013nice} uses low-level network switch properties to locate and map all the switches on a subnet and then associate rogue systems with specific physical switches. This is done automatically without relying on traditional network management tools and protocols (e.g., SNMP) which typically presume some prior knowledge of the network topology and often require administrative credentials. A number of agentless cloud computing endpoints monitoring has been proposed to support monitoring capability at the cloud to inspect and analyze endpoints attempting to connect to cloud services without having to stall the software on every user device~\cite{berlin2015malicious,brattstrom2017scalable}.
}


\vspace{1mm}
\noindent\textbf{Limitation of Anomaly Detection.} 
Although anomaly detection techniques are useful to discover novel and unknown attacks, there are still some challenges in terms of training and learning capabilities of those techniques. For instance, the network traffic is very complex and unpredictable especially in a dynamic environment. Thus, the model is subject to changes over time because anomalies are continuously evolving. Due to the changes in attack techniques and patterns, the information gained (trained) previously may be invalid.



\subsection{{Current Tools}}
{In this section, we provide an overview on the tools and prototypes which have already been used in design and development of SA monitoring systems. It includes the data collection, pre-processing, processing, and more comprehensive analysis which can be further used in SA systems. We discuss the existing SA related frameworks, prototypes and tools which have been implemented using research projects and real-world systems.
\subsubsection{Log file collector and analysis}
Event logging and network traffic analysis tools play a crucial role in designing and developing SA systems. In here, we summarize some of them used for SA systems. 
\begin{itemize}
\item[--] {\em SEC}: Simple Event Correlator (SEC) tool processes the text lines in the log files aiming to detect the certain event groups over the defined time window~\cite{vaarandi2005tools}. It can analyse and find the frequent patterns from the log files using the data mining algorithms such as breadth-first event log detection methods. SEC have been be used for designing SA monitoring systems in both data pre-processing and analysis phases~\cite{vaarandi2018unsupervised}. 
\item[--] {\em NTE}: Network Traffic Exploration (NTE)~\cite{vandenberghe2008network} is a security event packet analysis tool which can be use full to monitor the network traffic, analyse them, and detect various network attacks. It can be leveraged by SA monitoring systems tool either collect the network traffic information or detect the attacks using the pre-defined algorithms.
\item[--] {\em CogLog}: Cognitive Case Log (CogLog) The CogLog is a semantic tool which can keep a log of findings of the given investigation. CogLog has been used in the SA monitoring studies such as~\cite{bradshaw2012sol,bunch2012human}.
\item[--] {\em PANOPTESEC}: PANOPTESEC is a tool that manages the system's architecture and knowledge based on the security events and existing vulnerabilities. It can collect and further correlates the log files and alerts to detect attacks. The processed information can be returned to the users or system administrators. PANOPTESEC has been used in the SA monitoring system in the pre-processing and analysis phases~\cite{angelini2015percival,janiszewski2019novel}.
\item[--] {\em NECOMA}: This is a designed tool that collects the network traffic data from the network devices such as switches, routers, and IDS. It further analyzes the collected data to identify any attack attempts and mitigate the attacks~\cite{janiszewski2019novel}.
\end{itemize}
\subsubsection{Attack Graph generator tools}
\begin{itemize}
\item[--] {\em NetSPA:} Network Security Planning Architecture (NetSPA)~\cite{williams2008interactive} is an Attack Grapph (AG) generator and reachability analysis tool which consists of graphing subsystem component to visualize the computed attack graph. It can provide the assessment component of the survivable system. To address the scalability issue, it uses methods to prune the graph and make it simpler by removing the paths that do not reach the goal. NetSPA has been utilized by various studies for network monitoring and security evaluation systems such as~\cite{okhravi2011achieving}. 
\item[--] {\em GARNET:} GARNET~\cite{williams2008garnet} is an extended version of NetSpa which is able to capture the reachability of physical and logical topology by leveraging a graph subsystem based on tree maps. It is also able to evaluate the actual network situation through the interaction with the system~\cite{angelini2019mad}.
\end{itemize}
\subsubsection{Threat analysis tools}
The tools discussed before were limited to the capturing and analyzing network flow information and lacked the capability to monitor and detect threats and vulnerabilities~\cite{lakkaraju2004nvisionip, yin2005design}. However, various tools have been designed and developed to monitor and analyze the threats and obtain high level of situation awareness such as perception and projection~\cite{xi2011cnssa}. 
\begin{figure}[t]
	\centering
	\includegraphics[width=0.95\linewidth]{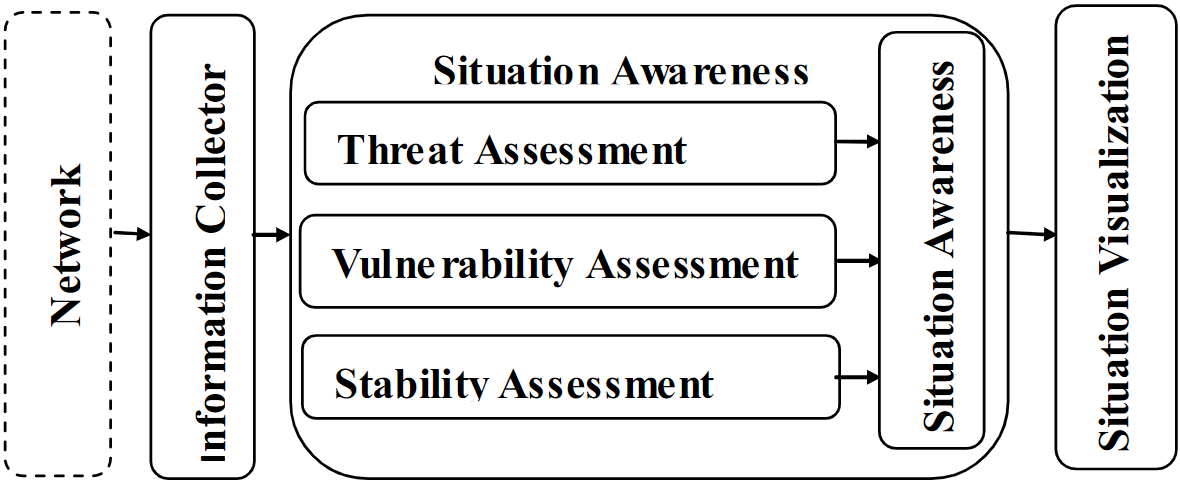}
	\caption{CNSSA framework developed by~\cite{xi2011cnssa}.}
	\label{fig:CNSSA}
\end{figure}
{
\begin{itemize}
\item[--] {\em CNSSA:} \citet{xi2011cnssa} developed a real-time situation awareness tool named Comprehensive Network Security Situation Awareness (CNSSA) which is enable to monitor the network environment based on the collected data and quantify the network situation awareness based on four metrics: Security, Threat, Vulnerability, and Stability. CNSSA architecture is presented in Fig.~\ref{fig:CNSSA}. CNSSA is equipped with a useful visualization and monitoring module which is able to illustrate network situation based on detailed multi-level view with various threat viewing features. 
\item[--] {\em Sol}: \citet{bradshaw2012sol} developed a cyber security situation awareness framework named as Sol. It analyze the cyber situation awareness using mutli-agent environment.
\item[--] {\em PERCIVAL}: \citet{angelini2015percival} proposed a novel visual analytics environment that obtains situational awareness providing the users with the understand of the network security posture and help them to monitor security events such as reactive and proactive attacks that are happening on the system.
\item[--] {\em MAD}: \citet{angelini2019mad} developed a Multi-step cyber Attack Detection (MAD) Visual Analytics solution aiming to improve the network security by analyzing the possible attacks and identifying suitable mitigation techniques.
\item[--] {\em Vulnus}: \citet{angelini2019mad} designed a visual analytics tool named 
VULNUS for dynamically inspecting the vulnerabilities spread on networks which helps to understand the network situation awareness. The proposed tool can visually classify nodes according to their vulnerabilities and compute the approximated optimal sequence of patches able to eliminate all the attack paths and allows for exploring sub-optimal patching strategies.
\end{itemize}
However, most of the existing tools cannot still capture more advanced threats mainly because of lack of appropriate (i) data collection module for collecting information related to advanced threats and (i) analysis module for evaluating the collected data and information to be able to discover advanced threats.
}}


\section{{Situation Awareness }} \label{sec:SA}
\subsection{Threat Evaluation}
Threat Evaluation in SA falls into a layer between cyber situation comprehension and projection, as it needs to provide higher perspective of the current situation of the threats based on its current perspective and its future impact. The more comprehension gained in this layer, the more SA level understanding could be obtained.
\subsubsection{Damage Assessment} The ability of being aware of the impact of the attack and threats, and vulnerability analysis is the main part of Damage Assessment (also called Impact Assessment)~\cite{liu2010cross}. The appropriate cyber situation awareness can help the security experts to make the right defense decisions and take select appropriate defense actions. The security analysts need to perform the three basic awareness stages (Situation Perception, Situation Comprehension, and Situation Projection) to gain enough cyber situational awareness under severe cyber attacks. Damage assessment is an essential component of the impact assessment and situation assessment in the situation comprehension stage and has been studies in various research~\cite{rajivan2017impact,liu2010cross}. Predictive damage assessment is an important part of situation projection which evaluate and analyze the damages which are going to be caused in (near) future which is missing in the current literature.





\subsubsection{Attack Tracking and Prediction} Network security situation evaluation methods based on attack intention recognition have been studied in literature~\cite{kou2019research,kun2015network,guang2016network}. 
\citet{kou2019research} proposed a method to recognize the attack intention on a network based on achieved attack phase and vulnerabilities in order to trace the next attack phase. They formulate the security situation ($Sa$) as Equation~\ref{eq:Sa}.
\begin{equation}\label{eq:Sa}
Sa=\sum_{i=1}^{n}{sa(path_i),}
\end{equation}
where, $Sa$ denotes the effect of each attack path on the network security situation which can be calculated based on the probability of attack stage with the attack threat and some weighted values.
Then, the predicted security situation (${Sa}'$) is defined as Equation~\ref{eq:Sa2}.
\begin{equation}\label{eq:Sa2}
{Sa}'=Sa+\sum_{i=1}^{l}{E_i(path),}
\end{equation}
where $l$ is the quantity of attack path, and $E_i$ is the effect of the attack on the next attack stage. The proposed method can evaluate the network security situation based on attack intention and stage recognition. This technique is used as situation projection to further predict the next attack stage. However, their proposed technique only works based on the known attacks and cannot be evaluated based on unknown or new-type attacks.

Furthermore, \citet{hu2017quantitative} proposed a comprehensive situation prediction model based on the overall network situation factors such as attacker, defender, and environment to show the adversary characteristic. Their proposed solution incorporates some important factors such as attack intention recognition, path detection, and success probability prediction. They further evaluated the threats using calculation of threat severity of critical assets and control the security situation. However, achieving those factors provide a high level of SA projection which are useful for further decision making and response. Similarly, various threat evaluation methods have been proposed for different purpose such as attack speed prediction~\cite{fredj2015realistic}, attack capacity inference~\cite{liu2016network}, attack goal identification~\cite{fredj2015realistic,liu2016network,nandi2016interdicting}, attack path prediction~\cite{nandi2016interdicting,fredj2015realistic,liu2016network}, success probability prediction~\cite{qu2010network,liu2016network}, and attack time prediction~\cite{hu2017quantitative}.

\subsection{Decision Making and Planning}
The SA system should be able to interact with situation response in which a planned course of actions needs to be taken~\cite{barford2010cyber}. Thus, before deploying any planned action, a decision making based on the consequence of the planned action should be done. {SA enables a decision maker's awareness of a situation and their understanding of the situation up to the point the decision is made. Once a decision is reached, planning and execution (of the response actions) occur.}

\begin{figure}[t]
	\centering
	\includegraphics[height=5.3cm]{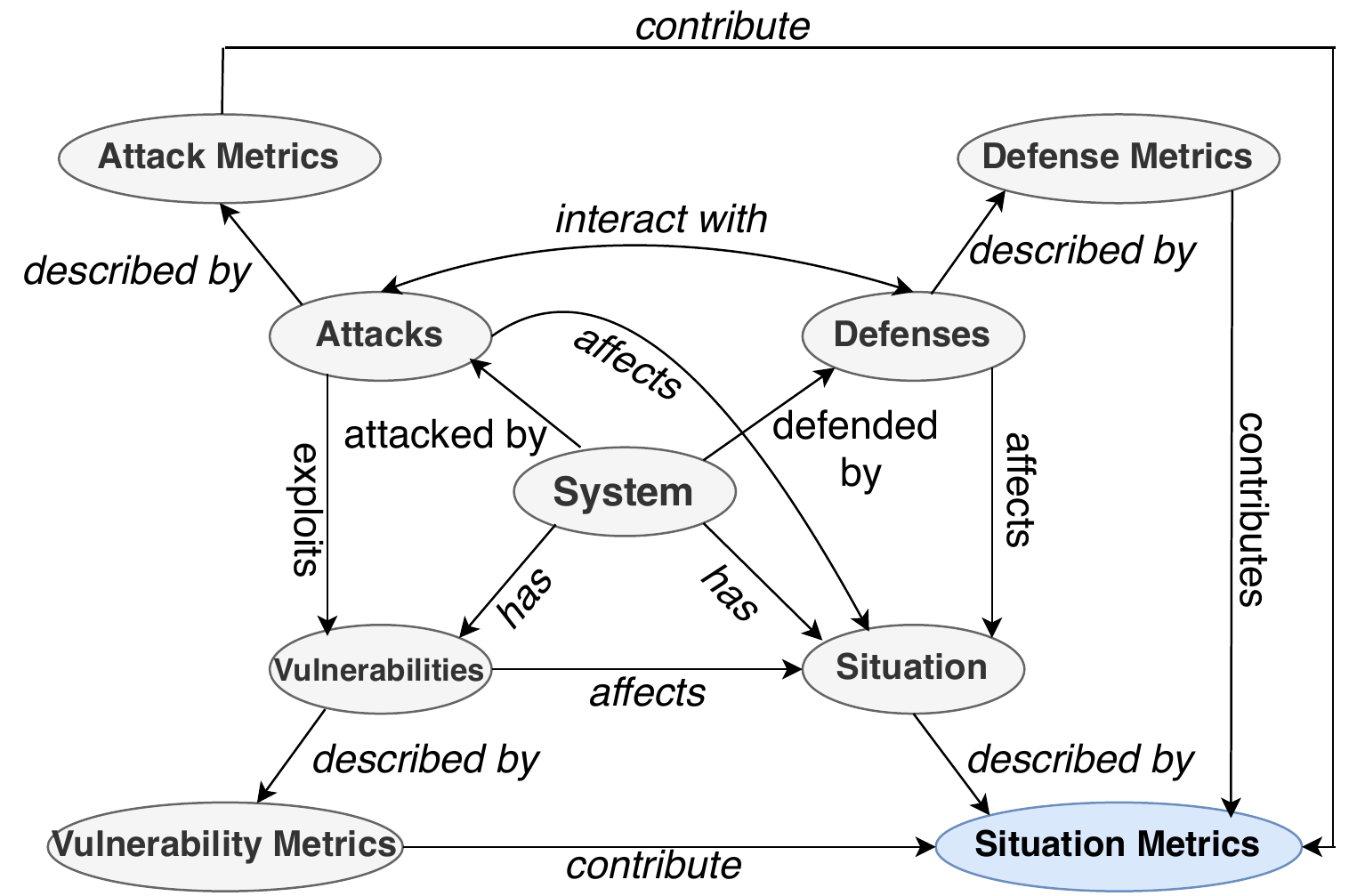}
	\caption{Decision making for SA based on vulnerabilities, attacks, and defense metrics~\cite{pendleton2016survey}.}
	\label{fig:SA-rel}
\end{figure}

Making an appropriate decision for a given system can be determined based on the system's situation. \citet{pendleton2016survey} defined the evolution of the situation for a given system based on a function of time which can be determined based on three attributes each of which can be represented as a function of time. They formulated the situation by defining the situation metrics by contributions of vulnerability, attack, and defense metrics, as shown in Equation~\ref{eq:S}. The relationship between those aspects is described as a model in Fig.~\ref{fig:SA-rel}.

\begin{equation}\label{eq:S}
situation(t)=f(V(t),D(t),A(t)),
\end{equation}
where $V(t)$ represents a function of existing vulnerabilities in the system at a given time $t$, and $D(t)$ and $A(t)$ denote the functions of defense and attack for time $t$. However, there still a preliminary progress towards explicit representation of the various kinds of required functions, $f$, corresponding to different situations and various attack-defense interactions.

Moreover, the decisions making in SA need to be evaluated based on the current and further security postures. For instance, the evaluation should be provided to show whether if the change in security practices cases negative impacts for further situations or not. This may include monetary costs, reputation damage, or so. However, it's important to be informed by an accurate understanding of the risks caused by a selected decision~\cite{webb2014situation}.


\subsection{Visualisation}

\begin{figure*}[t]
	\centering
	\includegraphics[height=6.9cm,width=0.79\textwidth]{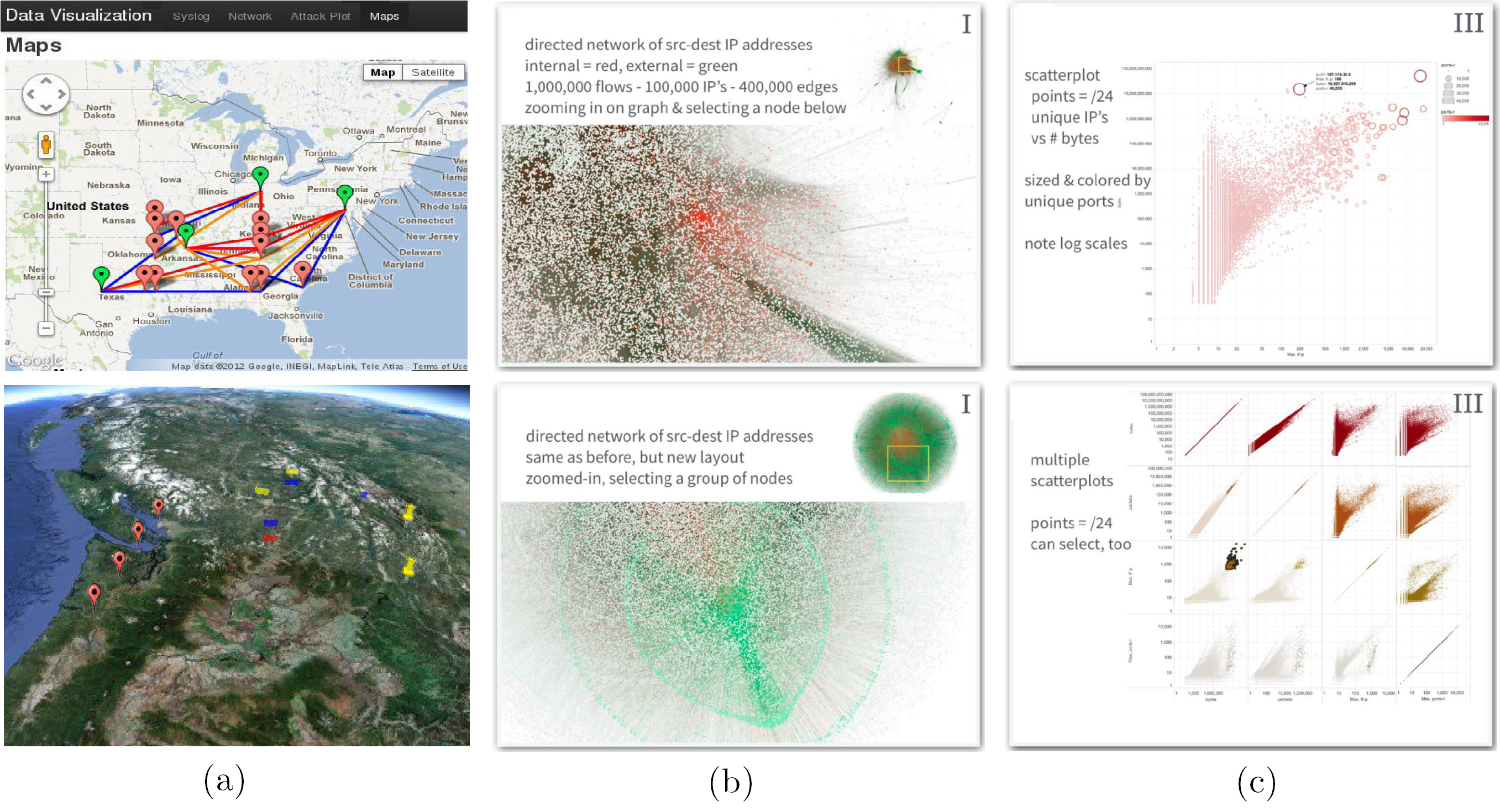}
	\caption{(a) Geo-location visualization of many-to-many attacks~\cite{yu2013detection}, (b) An overview of the network graph visualization used in the literature~\cite{mckenna2015unlocking}, (c) demonstration of chart-based visualization techniques.}
	\label{fig:visualization}
\end{figure*}

Visualization plays an important role in SA monitoring. It is a way to demonstrate the level of current threats, impacts, priorities, and sensitivity of analyzed data in the SA systems, and can be an interaction between computational process and human-based visual representation. This section discusses various visualization techniques used for SA such as statistical, historical, near real-time, and real-time presentations of SA system. 
Various existing SA tools use visualization techniques such as map-based, chart-based, network graph, line-charts, and flow diagrams to present information~\cite{healey2014visualizations,yu2013detection}. Visualization techniques for cybersecurity purposes can be demonstrated based on \textit{statistical summaries} which include some visualization techniques such as histogram, 2D or 3D graphs, and line-charts. Moreover, \textit{map-based} visualization techniques such as geo-locations views are useful tools to represent the cyber attack source and targets situations. Fig.~\ref{fig:visualization} shows some examples of this type of visualizations. \textit{Line-charts} visualization methods are useful for monitoring capabilities especially in real-time manner such as real-time sliding slice. In~\cite{fischer2014nstreamaware} real-time monitoring feature is used for visualizing Feature Selection.





Many studies have been proposed to develop the system design for analysis of data and visualize the threats aiming to support SA in real-time monitoring \cite{goodall2018situ,mansmann2012streamsqueeze}. For instance, \citet{mansmann2012streamsqueeze} proposed a screen-filling technique providing the events' details in a data stream in near real-time by following the history trends of the prior events. In another work, \citet{best2010real} utilized a behavioral modeling method which is able to learn the expected activities on a network. They presented a visualization system which combines various visualization techniques and tools, and provided the situational understanding of real-time network activities to help analysts to plan for further response steps.

\citet{healey2014visualizations} reviewed the scientific and information visualization of the proposed visualization systems for cyber SA. Then, they outlined a set of requirements to develop an appropriate visualization system for cyber SA domain. \citet{huffer2017situational} developed a proof-of-concept tool which is able to discover the roles of a system and helps cyber analysts to detect the changes in the network and devise a plan for incident response. Later on, in \cite{goodall2018situ}, they expanded their work by proposing an anomaly detection visualization system which can discover and explain suspicious activities and behaviors in the network's traffic and logs. They leveraged some visualization features such as a temporal histogram, horizon graphs, bar charts, and two-hop communication graph to demonstrate the network situation in real-time.

However, it is important to determine how best to leverage the visualization techniques for SA to allow analysts to monitor and detect unusual changes in the system, plan for incident response, and optimize the security posture.

\section{Discussion} \label{sec:discussion}
\subsection{Misconceptions}
A number of misconceptions around cyber SA realm have been discovered based on the current literature listed as follows:
\begin{itemize}
\item[--] A large volume of data collected about the system and threats is not considered as a part of situation awareness. For instance, intelligence reports such as threat and vulnerabilities. However, to make a monolithic SA system, all stages of gaining SA are required ranging from raw data, intelligence reports to high abstracted data.

\item[--] Intelligence and data sharing is not Cyber SA. Intelligence sharing data is still only one aspect of data gathering phase and can help the SA system in further stages to provide the system with better understanding of perceived or impending situation.

\item[--] However, threat intelligence reports solely should not be considered as cyber SA. Data and information gained through data gathering phase such as vulnerability and threats are only the source of SA analysis and situation awareness phases. Likewise, the outcome of intelligence reports alone needs to be used as an input for further analysis in the other phases and are not cyber SA.

\item[--] The large amount of collected data which is either raw, organized, or processed is not Cyber SA. This collected data can be useful in understanding partial situations. Thus, the collected big data is not SA, and only demonstrates one aspect and organized perspective of the situation and need to go through further analysis and evaluation.

\end{itemize}

\subsection{Insights and Limitations}
According to our extensive survey on the situation awareness systems, design, and development in the curret literature, we discovered that the following aspects of SA are still crude in current studies and will need to be further investigated.

\begin{itemize}
\item[--] More comprehensive data gathering including dealing with large amount of data, pre-processing, and parsing. Data collection need to be real-time which can help SA system to be updated based on the current situation of the system. However, using honeypots to collect and monitor the real data needs further investigation in the literature.

\item[--] Applying anomaly detection techniques in SA is still difficult for dynamic environment. Training and learning capabilities is challenging in this situation as the network traffic is very complex and unpredictable a dynamic environment. Thus, the model is subject to changes over time because anomalies are continuously evolving.

\item[--] There is still a lack of the comprehensive metric for SA which can quantify the system's current situation with the ability of capturing system's security/situation changes in real-time using SA-based metrics.

\item[--] There are still missing AI-based approaches for both attacker and defender sides such as modeling game and control theoretic approaches, adversarial modeling for AI-powered threats, artificial intelligence techniques, and human-computer interfaces. The vision of the future system is based on the side-by-side interaction between human analysts and the automated AI-based systems and tools. Moreover, the SA system design and development requires more sophisticated human-computer interaction and improvement on self-learning abilities for defensive and monitoring systems. This interaction will help the transaction from human-based defense system to AI-based systems in SA context more effectively and will enable the defenders to automatically prepare to defend against potential threats with quick adaption and automation capabilities to evolving cyber attacks.

\end{itemize}


\section{Conclusions}
The emerging threats are sophisticated, complex and highly dynamic and need to be addresses using situation awareness monitoring systems equipped with the ability to monitor and defend against wide range of attacks including AI-supported attacks. In this paper, we discussed the cyber SA taxonomy based on a comprehensive framework including different situation awareness levels such as data gathering mapped to situation perception, analysis and techniques mapped to situation comprehension, and finally situation awareness mapped to situation projection. We then conducted an extensive survey on each level of situation awareness category and discuss the current state-of-the-art for each and highlight the limitations. We also introduced the tools and prototypes which can be used for SA systems for either analysis or visualization phases.

\section*{Acknowledgement}
This work was supported by the Cyber Security Research Programme--''Artificial Intelligence for Automating Response to Threats'' from the Ministry of Business, Innovation, and Employment (MBIE) of New Zealand as a part of the Catalyst Strategy Funds under Grant MAUX1912.


\bibliographystyle{IEEETranSN}
\bibliography{SAM}


\end{document}